\documentclass[equations,11pt,showkeys]{article}
\usepackage{amssymb,amsmath,amsthm}
\usepackage{eucal}
\usepackage{eepic}
\usepackage{epsfig,epic,picinpar}
\usepackage{a4,tikz}
\usepackage{tikz}
\usepackage[all]{xy}
\usepackage{color}
\usepackage{accents}
\usepackage{texdraw}
\usepackage[toc]{appendix}

\begin{document}
\bibliographystyle{plain}
\def\m@th{\mathsurround=0pt}
\mathchardef\bracell="0365
\def\upbrall{$\m@th\bracell$}
\def\undertilde#1{\mathop{\vtop{\ialign{##\crcr
    $\hfil\displaystyle{#1}\hfil$\crcr
     \noalign
     {\kern1.5pt\nointerlineskip}
     \upbrall\crcr\noalign{\kern1pt
   }}}}\limits}
\mathchardef\hatcell="0362
\def\dobrall{$\m@th\hatcell$}
\def\underhat#1{\mathop{\vtop{\ialign{##\crcr
    $\hfil\displaystyle{#1}\hfil$\crcr
     \noalign
     {\kern1.5pt\nointerlineskip}
     \dobrall\crcr\noalign{\kern1pt
   }}}}\limits}
\usetikzlibrary{arrows,automata}
\usetikzlibrary{positioning}
\def\theequation{\arabic{section}.\arabic{equation}}
\newcommand{\aar}{\alpha}
\newcommand{\bb}{\beta}
\newcommand{\vth}{\vartheta}
\newcommand{\gm}{\gamma}
\newcommand{\Gm}{\Gamma}
\newcommand{\bs}{\boldsymbol}
\newcommand{\en}{\epsilon}
\newcommand{\ven}{\varepsilon}
\newcommand{\dd}{\delta}
\newcommand{\sg}{\sigma}
\newcommand{\kp}{\kappa}
\newcommand{\ld}{\lambda}
\newcommand{\oa}{\omega}
\newcommand{\half}{{\small \frac{1}{2}}}
\newcommand{\be}{\begin{equation}}
\newcommand{\ee}{\end{equation}}
\newcommand{\bea}{\begin{eqnarray}}
\newcommand{\eea}{\end{eqnarray}}
\newcommand{\bse}{\begin{subequations}}
\newcommand{\ese}{\end{subequations}}
\newcommand{\nn}{\nonumber}
\newcommand{\vf}{\varphi}
\newcommand{\sn}{{\rm sn}}
\newcommand{\cn}{{\rm cn}}
\newcommand{\dn}{{\rm dn}}
\newcommand{\wh}{\widehat}
\newcommand{\ol}{\overline}
\newcommand{\wt}{\widetilde}
\newcommand{\ut}{\undertilde}
\newcommand{\uh}{\underhat}
\newcommand{\ip}{{i^\prime}}
\newcommand{\jp}{{j^\prime}}
\newcommand{\btP}{\,^{t\!}{\bf P}}
\newcommand{\bI}{{\bf I}}
\newcommand{\bO}{{\bf O}}
\newcommand{\bA}{{\bf A}}
\newcommand{\bB}{{\bf B}}
\newcommand{\bU}{{\bf U}}
\newcommand{\bC}{{\bf C}}
\newcommand{\bOm}{{\bf \Omega}}
\newcommand{\buk}{{\bf u}_k}
\newcommand{\bul}{{\bf u}_\ell}
\newcommand{\tII}{\,^{t\!}{\bf I}}
\newcommand{\tuk}{\,^{t\!}{\bf u}_{k^\prime}}
\newcommand{\tul}{\,^{t\!}{\bf u}_{\ell^\prime}}
\newcommand{\tull}{\,^{t\!}{\bf u}_{-\ell+\ld}}
\newcommand{\tuq}{\,^{t\!}{\bf u}_{-q_j+\ld}}
\newcommand{\tck}{\,^{t\!}{\bf c}_{k^\prime}}
\newcommand{\ssk}{\sigma_{k^\prime}}
\newcommand{\ssl}{\sigma_{\ell^\prime}}
\newcommand{\pte}{(\partial_t-\partial_\eta)}
\newcommand{\pxe}{(\partial_x-\partial_\eta)}
\newcommand{\dint}{\int_\Gamma d\mu(\ell) }
\def\hypotilde#1#2{\vrule depth #1 pt width 0pt{\smash{{\mathop{#2}
\limits_{\displaystyle\widetilde{}}}}}}
\def\hypohat#1#2{\vrule depth #1 pt width 0pt{\smash{{\mathop{#2}
\limits_{\displaystyle\widehat{}}}}}}
\def\hypo#1#2{\vrule depth #1 pt width 0pt{\smash{{\mathop{#2}
\limits_{\displaystyle{}}}}}}

\newcommand{\pii}{P$_{{\rm\small II}}$}  
\newcommand{\pvi}{P$_{{\rm\small VI}}$}   
\newcommand{\ptf}{P$_{{\rm\small XXXIV}}$}   
\newcommand{\diffE}{O$\triangle$E}   

\newcommand{\ssa}{\frak{a}}
\newcommand{\ssb}{\frak{b}}
\newcommand{\ssc}{\frak{c}}
\newcommand{\ssp}{\frak{p}}
\newcommand{\ssq}{\frak{q}}
\newcommand{\ssr}{\frak{r}}

\newcommand{\pp}{\partial}
\newcommand{\hf}{\frac{1}{2}}
\newcommand{\ith}{$i^{\rm th}$\ }
\newcommand{\bu}{{\boldsymbol u}}
\newcommand{\bell}{{\boldsymbol l}}
\newcommand{\bj}{{\boldsymbol j}}
\newcommand{\bt}{{\boldsymbol t}}
\newcommand{\bm}{{\boldsymbol m}}
\newcommand{\boa}{{\boldsymbol \omega}}
\newcommand{\bet}{{\boldsymbol \eta}}
\newcommand{\bW}{\bar{W}}
 \newcommand{\pl}{\partial}
 \newcommand{\ddp}{\frac{\partial}{\partial p}}
 \newcommand{\ddq}{\frac{\partial}{\partial q}}
 \newcommand{\ddr}{\frac{\partial}{\partial r}}
 \newcommand{\Ld}{{\boldsymbol \Lambda}}
 \newcommand{\tLd}{\,^{t\!}{\boldsymbol \Lambda}}
 \newcommand{\I}{{\bf I}}
 \newcommand{\bP}{\boldsymbol{P}}
 \newcommand{\tbP}{\,^{t\!}\boldsymbol{P}}
 \newcommand{\tbC}{\,^{t\!}\boldsymbol{C}}
 \newcommand{\ddint}{\int_\Gamma d\ld(\ell) }
 \newcommand{\vE}{\vec{E} }
 \newcommand{\vL}{\vec{L} }
 \newcommand{\vn}{\vec{n} }
 \newcommand{\vR}{\vec{R} }
 \newcommand{\vP}{\vec{P} }
 \newcommand{\vna}{\vec{\nabla} }
 \newcommand{\vv}{\vec{v} }
 \newcommand{\vF}{\vec{F} }
 \newcommand{\vj}{\vec{j} }
 \newcommand{\vB}{\vec{B} }
 \newcommand{\vr}{\vec{r} }
 \newcommand{\vp}{\vec{p} }
 \newcommand{\vk}{\vec{k} }
\newcommand{\mbe}{{\boldsymbol e}}
\newcommand{\bE}{{\boldsymbol E}}
\newcommand{\bV}{{\boldsymbol V}}
\newcommand{\bnab}{{\boldsymbol \nabla}}
\newcommand{\buu}{{\boldsymbol u}}
\newcommand{\bv}{{\boldsymbol v}}
\newcommand{\ba}{{\boldsymbol a}}
\newcommand{\bbb}{{\boldsymbol b}}
\newcommand{\bS}{{\boldsymbol S}}
\newcommand{\bT}{{\boldsymbol T}}
\newcommand{\bJ}{{\boldsymbol J}}
\newcommand{\bc}{{\boldsymbol c}}
\newcommand{\bw}{{\boldsymbol w}}
\newcommand{\mbx}{{\boldsymbol x}}
\newcommand{\mby}{{\boldsymbol y}}
\newcommand{\bz}{{\boldsymbol z}}
\newcommand{\brr}{{\boldsymbol r}}
\newcommand{\bp}{{\boldsymbol p}}
\newcommand{\bk}{{\boldsymbol k}}
\newcommand{\btt}{{\boldsymbol t}}
\newcommand{\bmm}{{\boldsymbol m}}
\newcommand{\bdd}{{\boldsymbol \delta}}
\newcommand{\bze}{{\boldsymbol 0}}
\newcommand{\boma}{{\boldsymbol \omega}}
\newcommand{\bxi}{{\boldsymbol \xi}}
 \newcommand{\mbv}{\boldmath{v}}
 \newcommand{\mbxi}{\boldmath{\xi}}
 \newcommand{\mbeta}{\boldmath{\eta}}
 \newcommand{\mbw}{\boldmath{w}}
 \newcommand{\mbu}{\boldmath{u}}

 \def\hypotilde#1#2{\vrule depth #1 pt width 0pt{\smash{{\mathop{#2}
 \limits_{\displaystyle\widetilde{}}}}}}
 \def\hypohat#1#2{\vrule depth #1 pt width 0pt{\smash{{\mathop{#2}
 \limits_{\displaystyle\widehat{}}}}}}
 \def\hypo#1#2{\vrule depth #1 pt width 0pt{\smash{{\mathop{#2}
 \limits_{\displaystyle{}}}}}}

\newtheorem{lemma}{Lemma}[section]
\newtheorem{cor}{Corollary}[section]
\newtheorem{prop}{Proposition}[section]
\newtheorem{definition}{Definition}[section]
\newtheorem{conj}{Conjecture}[section]
\newtheorem*{theorem}{Theorem}

\newtheoremstyle{named}{}{}{\itshape}{}{\bfseries}{.}{.5em}{\thmnote{#3's }#1}
\theoremstyle{named}
\newtheorem*{namedtheorem}{Theorem}

\begin{flushright}
\end{flushright}
\begin{center}
{\large{\bf On Elliptic Lax Systems on the Lattice and a Compound Theorem for Hyperdeterminants}}
\vspace{.4cm}

N. Delice$^1$, F.W. Nijhoff$^1$ and S. Yoo-Kong$^2$ \\
{\it $^1$Department of Applied Mathematics, School of Mathematics, University of Leeds,\\ United Kingdom, LS2 9JT.}\\
{\it $^2$Theoretical and Computational Physics (TCP) Group, Department of Physics,\\ Faculty of Science, King Mongkut's University of Technology Thonburi,\\ Thailand, 10140.}
\vspace{.2cm}

\end{center}

\vspace{.4cm}
\centerline{\bf Abstract}
\vspace{.2cm}

\noindent
A general elliptic $N\times N$ matrix Lax scheme is presented, leading to two classes of elliptic lattice systems, one 
which we interpret as the higher-rank analogue of the Landau-Lifschitz equations, while the other class we characterize as 
the higher-rank analogue of the lattice Krichever-Novikov equation (or Adler's lattice). We present the general scheme, 
but focus mainly of the latter type of models. In the case $N=2$ we obtain a novel Lax representation of 
Adler's elliptic lattice equation in its so-called 3-leg form. The case of rank $N=3$ is analysed using Cayley's hyperdeterminant 
of format  $2\times2\times2$, yielding a multi-component system of coupled 3-leg quad-equations.

\pagebreak



\newpage

\section{Introduction}
\setcounter{equation}{0}
\label{sec:intro}

Adler's lattice equation, \cite{Adler2}, is an integrable lattice version of the Krichever-Novikov (KN) equation, \cite{KN}, i.e. of the nonlinear evolution equation
\begin{eqnarray}\label{eq:KN}
u_t = \frac{1}{4}\left( u_{xxx}+\frac{3}{2}\,\frac{r(u)-u^2_{xx}}{u_x}\,\right)\  ,
\end{eqnarray}
in which $r(u) = 4u^3-g_2u-g_3$ is the polynomial associated with a Weierstrass elliptic curve (or more generally an arbitrary
quartic polynomial). This lattice equation, which was obtained as the permutability condition for the B\"acklund transformations
for \eqref{eq:KN}, can be written in the form\footnote{Note that in the original paper \cite{Adler2} the equation was written in a
slightly different form with rather complicated expressions for the coefficients given in terms of the moduli $g_2$ and $g_3$ of the
Weierstrass curve.}:
\begin{eqnarray}
&&A\left[ (u-b)(\wh{u}-b)-(a-b)(c-b)\right]
\left[(\wt{u}-b)(\wh{\wt{u}}-b) -(a-b)(c-b)\right] \nn \\
&&+B\left[ (u-a)(\wt{u}-a)-(b-a)(c-a)\right]
\left[ (\wh{u}-a)(\wh{\wt{u}}-a) -(b-a)(c-a)\right]= \nn \\
&& = ABC(a-b)\  ,    \label{eq:ellform}
\end{eqnarray}
cf. \cite{Nij}, where $u=u(n,m)$ is the dependent variable, with the shifted variables $\wt{u}=u(n+1,m)$, $\wh{u}=u(n,m+1)$ and
$\wh{\wt{u}}=u(n+1,m+1)$ defining the different values of $u$ at the vertices around an elementary plaquette, cf. Figure $\ref{EPlaq}$.
The $\ssa$,$\ssb$ in Figure $\ref{EPlaq}$ are lattice parameters associated with the grid size, and in this elliptic equation 
they are points $\ssa=(a,A)$, $\ssb=(b,B)$, together with $\ssc=(c,C)$, on a Weierstrass elliptic curve, i.e.
\begin{equation}\label{aA}
A^2=r(a)\equiv 4a^3-g_2a-g_3 ~~~~,~~~~ B^2=r(b)~~~~,~~~~ C^2=r(c)\  ,
\end{equation}
which can parametrised in terms of the Weierstrass $\wp-$function as follows:
\begin{align}\label{eq:ABC}
(a,A)=(\wp(\alpha),\wp'(\alpha)),\hspace{5 mm} (b,B)=(\wp(\beta),\wp'(\beta)),\hspace{5 mm} (c,C)=(\wp(\gamma),\wp'(\gamma))\ ,
\end{align}
where $\alpha$ and $\beta$ are the corresponding uniformising parameters and where $\gamma=\beta-\alpha$. 
The parameters $\ssa$, $\ssb$ and $\ssc$ are related through the addition formulae on the elliptic curve:
\begin{eqnarray}\label{addition1}
A(c-b)&=&C(a-b)-B(c-a),\nn \\
a+b+c&=&\frac{1}{4}\left(\frac{A+B}{a-b}\right)^2\  .
\end{eqnarray}
Furthermore, we use the notation for the lattice shifts
$$
 u\overset{\alpha }{\longrightarrow}\widetilde{u}\  , \qquad u\overset{\beta}{\longrightarrow}\widehat{u}
$$
being the elementary shifts on a quadrilateral lattice, each being associated with the lattice parameters $(a,A)$ respectively $(b,B)$, with the equation \eqref{eq:ellform}
expressing the condition for commutativity of these shifts as expressed through the diagram:
\vspace{.3cm}

\begin{figure}[ht]
\centering
\begin{tikzpicture}[every node/.style={minimum size=1cm},on grid]
 \filldraw[fill=white!20] (0,0) rectangle (2,2);
 \fill (2,2) circle (0.1) node[above right=-0.3] {$\widetilde{u}$};
 \fill (0,0) circle (0.1) node[below left=-0.3] {$\widehat{u}$};
 \fill (0,2) circle (0.1) node[above left=-0.3] {$u$};
 \fill (2,0) circle (0.1) node[below right=-0.4] {$\widehat{\widetilde{u}}$};
 \node at (1,2.2) {$\ssa$};
 \node at (-0.3,1) {$\ssb$};
 \node at (2.2,1) {$\ssb$};
 \node at (1,-0.3) {$\ssa$};
\end{tikzpicture}
\caption{Configuration of lattice points in the lattice equation $(\ref{eq:ellform})$.}
\label{EPlaq}
\end{figure}
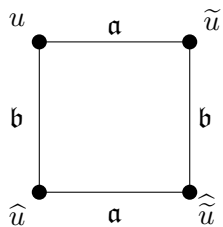
\vspace{.3cm}

A Lax pair for Adler's equation was given in \cite{Nij}, and the equation reemerged in \cite{ABS} as the top equation in the ABS list of affine-linear quadrilateral equations, where it
was renamed Q4. The key integrability characteristic of Adler's equation is its \textit{multidimensional consistency}, \cite{NW,BS}, which in the case of Adler's equation can be made
manifest through its so-called 3-leg form, cf. \cite{ABS}:
\begin{equation}
\label{eq:3leg}
\frac{\sg(\wt{\xi}-\xi+\alpha)\,\sg(\wt{\xi}+\xi-\alpha)}
{\sg(\wt{\xi}-\xi-\alpha)\,\sg(\wt{\xi}+\xi+\alpha)}\,
\frac{\sg(\wh{\xi}-\xi-\bb)\,\sg(\wh{\xi}+\xi+\bb)}
{\sg(\wh{\xi}-\xi+\bb)\,\sg(\wh{\xi}+\xi-\bb)}
=\frac{\sg(\wh{\wt{\xi}}-\xi-\gm)\,\sg(\wh{\wt{\xi}}+\xi+\gm)}
{\sg(\wh{\wt{\xi}}-\xi+\gm)\,\sg(\wh{\wt{\xi}}+\xi-\gm)}
\end{equation}
in which the uniformising variable $\xi=\xi(n,m)$ is now the dependent variable of the 
equation, related to the original variable $u$ of the rational form \eqref{eq:ellform} of the equation through the
identification $u=\wp(\xi)$. The connection between rational and elliptic form of the equation parallels that of the KN equation, 
which in its (original) elliptic form reads:
\begin{eqnarray}\label{eq:ellKN}
\xi_t = \frac{1}{4}\left( \xi_{xxx}+\frac{3}{2}\,\frac{1-\xi^2_{xx}}{\xi_x}-6\wp(2\xi)\,\xi_x^3\,\right)\  .
\end{eqnarray}

We note in passing that there are alternative forms for Adler's equation based on different choices of the underlying elliptic 
curve. Thus, if one could consider \eqref{eq:ellform} to be
the Weierstrass form of the equation (with parameters on a Weierstrass elliptic curve \eqref{aA}), the equation in Jacobi form 
(due to Hietarinta, \cite{H}) reads:
\begin{eqnarray}\label{Jacobi}
Q(v,\widetilde{v},\widehat{v},\widehat{\widetilde{v}})=p(v\widetilde{v}+\widehat{v}\widehat{\widetilde{v}})-q(v\widehat{v}+\widetilde{v}\widehat{\widetilde{v}})
-r(\widetilde{v}\widehat{v}+v\widehat{\widetilde{v}})+pqr(1+v\widetilde{v}\widehat{v}\widehat{\widetilde{v}})=0
\end{eqnarray}
where the dependent variable $v$ is related to $u$ of \eqref{eq:ellform} through a fractional linear transformation, and where the parameters $(p,P)$,\, $(q,Q)$ and $(r,R)$ are now
points on a Jacobi type elliptic curve:
\begin{equation}\label{pP}
\Gamma: \quad  X^2 \equiv x^4-\gamma x^2 +1, \quad \quad \gamma^2=k+1/k,\;
\end{equation}
with modulus $k$. They can be parametrised in terms of Jacobi elliptic function as follows:
\begin{align}\label{PQR}
&\ssp=(p,P)=( \sqrt{k}\;\rm sn(\alpha;k),\rm sn'(\alpha;k)),\hspace{5 mm} \ssq =(q,Q)=(\sqrt{k}\;\rm sn(\beta;k),\rm sn'(\beta;k)),\nn\\
&\ssr=(r,R)=(\sqrt{k}\;\rm sn(\alpha-\beta;k),\rm sn'(\alpha-\beta;k))\  .
\end{align}
Many interesting results were established for the latter form of the equation, notably explicit expressions for the (doubly elliptic) $N$-soliton solutions, \cite{AN},
however for the sake of the present paper we will concentrate once again on the Weierstrass form of the equation.

In the present paper we propose a general elliptic Lax scheme of rank $N$, which is inspired by a novel Lax representation 
of Adler's lattice equation. This Lax scheme leads to two distinct classes of systems which we coin as being 
"of Landau-Lifschitz type" (or spin-nonzero case) and as "of Krichever-Novikov type" (or spin-zero case). 
We present general results for both classes in section 2, but then focus in the remainder of the paper on the 
Krichever-Novikov class of Lax systems.  In that case for  $N=2$ we show that the scheme amounts to a novel Lax representation 
for Adler's lattice equation, which yields the equation directly in 3-leg form (this in contrast with the lax pair constructed in 
\cite{Nij} from multidimensional consistency). Notably in the rank $N=3$ case the analysis of the compatibility condition
exploits a (to our knowledge novel) \textit{compound theorem} for Caley's hyperdeterminants of format $2\times2\times2$, 
\cite{Cayley}, a result which may have some significance in its own right. We conjecture that the resulting rank 3 lattice 
system may be regarded as a discrete analogue of a rank 3 Krichever-Novikov type of differential system that was constructed
by Mokhov in \cite{Mokh1}.

\section{General Elliptic Lax Scheme}
\setcounter{equation}{0}

Consider the Lax pair of the form:
\bse \label{eq:RLax} \begin{eqnarray}
\wt{\chi}_\kp&=&L_\kp\,\chi_\kp\  , \label{eq:RLaxa}  \\
\wh{\chi}_\kp&=&M_\kp\,\chi_\kp\  . \label{eq:RLaxb} 
\end{eqnarray}\ese
defining horizontal and vertical shifts of the vector function $\chi_\kp$, according to the diagram:
\begin{figure}[ht]
\centering
\begin{tikzpicture}[every node/.style={minimum size=1cm},on grid]
\draw[thick,dashed,->] (0,0) -- (2,0) node[above right=-0.3] {$\widehat{\widetilde{\chi}}$};
\draw[thick,dashed,->] (0,2) -- (0,0) node[above left=-0.3] {$\widehat{\chi}$};
\draw[thick,<-] (2,0) -- (2,2) node[above right=-0.3] {$\widetilde{\chi}$};
\draw[thick,<-] (2,2) -- (0,2) node[below=-0.8] {$\chi$};
\node at (1,2.3) {$L$};
 \node at (-0.5,1.1) {$M$};
 \node at (2.3,1.1) {$\widetilde{M}$};
 \node at (1,-0.3) {$\widehat{L}$};
\end{tikzpicture}
\caption{Lax compatibility condition \eqref{eq:zerocurv}.}
\end{figure}
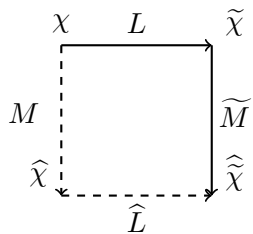
where the vectors $\chi$ are located at the vertices of the quadrilateral and in which the matrices L and M are attached to the edges linking the vertices. 
The matrices $L_\kp$ and$M_\kp$  can be taken of the form;
\bse \label{eq:Laxmast} \bea
 (L_\kp)_{i,j} &=& \Phi_{N\kp}(\wt{\xi}_i-\xi_j-\alpha) h_j\   ,  \\
 (M_\kp)_{i,j} &=& \Phi_{N\kp}(\wh{\xi}_i-\xi_j-\bb) k_j\   , \\
 && \qquad \qquad  (i,j=1,\dots,N)  \nn
\eea\ese
in which $\Phi_\kp$ denotes the (truncated) Lam\'e function
\begin{equation}\label{eq:Lame}
\Phi_\kp(\xi)\equiv \frac{\sg(\xi+\kp)}{\sg(\xi)\sg(\kp)}
\end{equation}
with $\sg$ denoting the Weierstrass $\sg$-function and the variables $\xi_i=\xi_i(n,m)$, ($i=1,\dots,N$), are the main dependent
variables. As before $\alpha$ and $\bb$ denote the uniformised lattice parameters (as in \eqref{eq:ABC}), while $\kp$ is the
(uniformised) spectral parameter.  In \eqref{eq:Laxmast}, the coefficients $h_j$, $k_j$, are some functions of the variables
$\xi_l$, and of their shifts, that remain to be determined.
The compatibility conditions between \eqref{eq:RLaxa} and \eqref{eq:RLaxb} are given by the lattice zero-curvature condition:
\be\label{eq:zerocurv} \wh{L}_\kp M_\kp=\wt{M}_\kp L_\kp\   . \ee
Using the addition formula
\be\label{eq:addform} \Phi_\kp(x)\Phi_\kp(y)=\Phi_\kp(x+y)\left[ \zeta(\kp)+\zeta(x)+\zeta(y)-\zeta(\kp+x+y)\right]\   , \ee
the consistency gives rise to
\begin{eqnarray}
&&  \sum_{l=1}^N \,\wh{h}_l k_j \left[\zeta(\wh{\wt{\xi}}_i-\wh{\xi}_l-\alpha)+\zeta(\wh{\xi}_l-\xi_j-\bb)
+ \zeta(N\kp)-\zeta(N\kp+\wh{\wt{\xi}}_i-\xi_j-\alpha-\bb)\right] = \nn \\
&& =\quad  \sum_{l=1}^N \,\wt{k}_l h_j \left[\zeta(\wh{\wt{\xi}}_i-\wt{\xi}_l-\bb)+\zeta(\wt{\xi}_l-\xi_j-\alpha)
+ \zeta(N\kp)-\zeta(N\kp+\wh{\wt{\xi}}_i-\xi_j-\alpha-\bb)\right] \nn  \\
&& \qquad \qquad\qquad  (i,j=1,\dots,N)\  .  \label{eq:compeqs}
\label{eq:compeqs}
\end{eqnarray}
Due to the arbitrariness of the spectral parameter $\kp$ the equations \eqref{eq:compeqs} separate into two parts, namely
\bse\label{sepeqs}\begin{eqnarray}
&& \left( \sum_{l=1}^N \wh{h}_l\right)k_j= \left(\sum_{l=1}^N \wt{k}_l \right) h_j\quad,\quad
(j=1,\dots,N)\  , \label{eq:sepeqsa} \\
&&  \left\{\sum_{l=1}^N \,\wh{h}_l\left[\zeta(\wh{\wt{\xi}}_i-\wh{\xi}_l-\alpha)+\zeta(\wh{\xi}_l-\xi_j-\bb) \right]\right\} k_j
= \left\{\sum_{l=1}^N \,\wt{k}_l\left[\zeta(\wh{\wt{\xi}}_i-\wt{\xi}_l-\bb)+\zeta(\wt{\xi}_l-\xi_j-\alpha)\right]\right\} h_j \nn  \\
&& \qquad\qquad\qquad (i,j=1,\dots,N)\  . \label{eq:sepeqsb}
\end{eqnarray}\ese

Now there are two scenarios which we refer to as the ``Landau-Lifschitz type'' (or physically, the spin non-zero) case and the
``Krichever-Novikov type'' (spin zero) cases respectively:
\begin{enumerate}
\item \underline{Discrete Landau-Lifschitz (LL) type case}: $\sum_l h_l\neq 0$, in which case we
have that the variables $h_j$, $k_j$ are proportional to each other, ~$k_j=\rho h_j$~, and after summing \eqref{eq:sepeqsa}
we obtain the conservation law:
\begin{equation}\label{eq:conserv}
\frac{\sum_{l=1}^N \wh{h}_l}{\sum_{l=1}^N h_l}=\frac{\sum_{l=1}^N \wt{k}_l}{\sum_{l=1}^N k_l}\  .
\end{equation}
and in which case eqs. (\ref{eq:sepeqsb}) reduce to:
\begin{eqnarray}
&&\sum_{l=1}^N \,\left[ \zeta(\wh{\wt{\xi}}_i-\wh{\xi}_l-\alpha) \rho\wh{h}_l-\zeta(\wh{\wt{\xi}}_i-\wt{\xi}_l-\bb) \wt{k}_l \right] =
\sum_{l=1}^N \,\left[ \zeta(\xi_j-\wh{\xi}_l+\bb) \rho\wh{h}_l -\zeta(\xi_j-\wt{\xi}_l+\alpha) \wt{k}_l \right]\  . \nn \\
&& \qquad\qquad\qquad (i,j=1,\dots,N)\  . \label{eq:zetasystem}
\end{eqnarray}
This system of equations can be reduced under the additional assumption of the conservation law (for the centre of mass):
\be\label{eq:Xi}
\wt{\Xi}+\wh{\Xi}=\wh{\wt{\Xi}}+\Xi\quad,\quad \Xi\equiv \sum_{l=1}^N \xi_l\   .
\ee
\item \underline{Krichever-Novikov (KN) type case}: $\sum_l h_l=\sum_l k_l=0$, in which case \eqref{eq:sepeqsa} becomes vacuous. 
In this case we seek further reductions by the additional constraint ~$\Xi=\sum_l \xi_l=0$ (modulo the period lattice of the elliptic functions).  
\end{enumerate}

In this paper we will focus primarily on the class of models in \# 2, but we will conclude this section by presenting the general 
structure of the systems that emerge from the Lax system in both cases, and then in the ensuing sections present an 
alternative analysis for the Lax system of class \# 2 for the cases $N=2$ and $N=3$. 

In order to proceed with the general analysis of \eqref{eq:zetasystem} we use a trick that was employed in \cite{NRK},
based on an elliptic version of the Lagrange interpolation formula (cf. Appendix \ref{appendix:b})  
in order to identify the variables $h_l$, $k_l$. 
Consider the following elliptic function, where as a consequence of the conservation law \eqref{eq:Xi} for the 
variables $\xi_l$ the Lagrange interpolation \eqref{eq:Lagr2} of Appendix \ref{appendix:b} 
is applicable, leading to the following identity:
\begin{eqnarray}\label{eq:F}
F(\xi)&=&\prod_{l=1}^N\,\frac{\sg(\xi-\wh{\wt{\xi}}_l)\sg(\xi-\xi_l-\alpha-\bb)}{\sg(\xi-\wh{\xi}_l-\alpha)\sg(\xi-\wt{\xi}_l-\bb)} 
\nn \\
&=& \sum_{l=1}^N \left[ \zeta(\xi-\wh{\xi}_l-\alpha)- \zeta(\eta-\wh{\xi}_l-\alpha)\right] H_l \nn \\
&& + \sum_{l=1}^N \left[ \zeta(\xi-\wt{\xi}_l-\bb)- \zeta(\eta-\wt{\xi}_l-\bb)\right] K_l
\end{eqnarray}
which holds for any four sets of variables $\xi_l$, $\wh{\xi}_l$, $\wt{\xi}_l$, $\wh{\wt{\xi}}_l$ such that \eqref{eq:Xi} holds.
In \eqref{eq:F} $\eta$ can be any one of the zeroes of $F(\xi)$, i.e. $\wh{\wt{\xi}}_i$ or
$\xi_i+\alpha+\bb$, and the coefficients $H_j$, $K_j$ are  given by:
\bse\begin{eqnarray}\label{eq:hk}
H_l&=& \frac{\prod_{k=1}^N \sg(\wh{\xi}_l-\wh{\wt{\xi}}_k+\alpha)\sg(\wh{\xi}_l-\xi_k-\bb)}
{\left[\prod_{k=1}^N \sg(\wh{\xi}_l-\wt{\xi}_k-\gm)\right]\prod_{k\neq l} \sg(\wh{\xi}_l-\wh{\xi}_k)} \\
K_l&=& \frac{\prod_{k=1}^N \sg(\wt{\xi}_l-\wh{\wt{\xi}}_k+\bb)\sg(\wt{\xi}_l-\xi_k-\alpha)}
{\left[\prod_{k=1}^N \sg(\wt{\xi}_l-\wh{\xi}_k+\gm)\right]\prod_{k\neq l} \sg(\wt{\xi}_l-\wt{\xi}_k)} \  .
\end{eqnarray}\ese
Furthermore, the coefficients obey the \textit{identity}:
\be\label{eq:HKidentity} \sum_{l=1}^N(H_l+K_l)=0\  . \ee
Taking $\xi=\wh{\wt{\xi}}_i$, $\eta=\xi_j+\alpha+\bb$ in (\ref{eq:F}) and comparing with
\eqref{eq:sepeqsb}, we can make the identifications:
\begin{equation}\label{eq:HK}
t H_l=\rho\wh{h}_l\quad ,\quad tK_l=-\wt{\rho}\wt{h}_l\quad ,\quad l=1,\dots, N\  ,
\end{equation}
with a function $t$ being an arbitrary proportionality factor. Thus in this case (case 1) by eliminating $h_l$ from \eqref{eq:HK}
we get the set of equations
\be\label{eq:LLeqs}
\frac{\wt{t}}{\wt{\rho}}\wt{H}_l+\frac{\wh{t}}{\wh{\wt{\rho}}} \wh{K}_l=0\quad ,\quad l=1,\dots,N
\ee
which, by inserting the expressions \eqref{eq:hk} for $H_l$ and $K_l$, is a system of $N$ equations for $N+2$
unknowns $\xi_l$, ($l=1,\dots,N$), and $\rho$ and $t$. Rewriting this system in explicit form, we obtain the system
of $N$ 7-point equations:
\be\label{eq:dLL}
\prod_{k=1}^N\,\frac{\sg(\xi_l-\wt{\xi}_k+\alpha)\,\sg(\xi_l-{\uh{\xi}}\!\!\!\phantom{a}_k-\beta)\,\sg(\xi_l-{\ut{\wh{\xi}}}\!\!\!\phantom{a}_k+\gamma)}
{\sg(\xi_l-\wh{\xi}_k+\beta)\,\sg(\xi_l-{\ut{\xi}}\!\!\!\phantom{a}_k-\alpha)\,\sg(\xi_l-{\uh{\wt{\xi}}}\!\!\!\phantom{a}_k-\gamma)}=-p
\ee
for $N+1$ variables $\xi_i$ ($i=1,\dots,N$) and $p=\ut{t}\,\uh{\rho}/(\uh{t}\,\rho)$, supplemented
with \eqref{eq:Xi} which fixes the discrete dynamics of the centre of mass $\Xi$ . In \eqref{eq:dLL} the under-accents 
$\ut{\cdot}$ and $\uh{\cdot}$ denote reverse lattice shifts, i.e., ${\ut{\xi}}\!\!\!\phantom{a}_i(n,m)=\xi_i(n-1,m)$ and 
${\uh{\xi}}\!\!\!\phantom{a}_i(n,m)=\xi_i(n,m-1)$ respectively.  These equations and their rational forms will be investigated 
more in detail in a future publication. We mention here only that the one-step periodic reduction, ~$\wt{\chi}_\kappa=\lambda\chi_\kappa$~, 
in this case leads to an implicit system of ordinary difference equations which amounts to a the time-discretization of the 
Ruijsenaars (relativistic Calogero-Moser) model, cf. \cite{NRK}.  In the remainder of the paper we will concentrate on the case \#2 which constitutes 
higher rank analogues of Adler's lattice equation in 3-leg form, and we will perform a different kind of analysis in that case.

\section{Elliptic Lax pairs for 3-leg lattice systems}
\setcounter{equation}{0}

In this section we will focus on case \#2 of general elliptic Lax systems introduced in the previous section, corresponding to 
the "spin-zero" case (where $\sum_{l=1}^N h_l=\sum_{l=1}^N k_l=0$). We will first demonstrate in the case $N=2$ of this system 
how the 3-leg form of Adler's equation arises in a natural way from this Lax pair. In fact, it turns out that the elaboration 
of the compatibility conditions for this Lax pair immediately produces the required equations, and is far less laborious than 
of the consistency-around-the-cube (CAC) Lax pair of \cite{Nij} yielding the corresponding rational form of Q4. Next we will analyse the much more generic 
case of $N=3$, and produce a novel system of elliptic lattice equations, which constitutes the main result of this paper. 
We also present the structure of the lattice system arising form the scheme for general $N$, based on similar ingredients as 
the ones used in the case \#1 elaborated in the previous section, but subject to slightly different conditions. 

\subsection{Case N=2: Elliptic Lax Pair for the Adler 3-leg lattice equation}
\setcounter{equation}{0}
Let $\xi=\xi_{n,m}$ be a function of the discrete independent variables $n$, $m$ for which we want
to derive a lattice equation from the following Lax pair:
\bse\label{eq:EllLax}\begin{eqnarray}
\wt{\chi}=L_\kp\chi&=&\ld\left( \begin{array}{cc}
\Phi_{2\kp}(\wt{\xi}-\xi-\alpha)&-\Phi_{2\kp}(\wt{\xi}+\xi-\alpha) \\
\Phi_{2\kp}(-\wt{\xi}-\xi-\alpha)&-\Phi_{2\kp}(-\wt{\xi}+\xi-\alpha) \end{array}
\right)\chi\\
\wh{\chi}=M_\kp\chi&=&\mu\left( \begin{array}{cc}
\Phi_{2\kp}(\wh{\xi}-\xi-\bb)&-\Phi_{2\kp}(\wh{\xi}+\xi-\bb) \\
\Phi_{2\kp}(-\wh{\xi}-\xi-\bb)&-\Phi_{2\kp}(-\wh{\xi}+\xi-\bb)\end{array}
\right)\chi\   ,
\end{eqnarray}\ese
in which the coefficients $\ld$ and $\mu$ are functions $\ld=\ld(\xi,\wt{\xi};\alpha)$ and
$\mu=\mu(\xi,\wh{\xi};\bb)$, respectively. The explicit form of which will be derived subsequently, but these forms will
actually not be relevant for the determination of the resulting lattice equation, which is Adler's system in 3-leg form.
The discrete zero-curvature condition \eqref{eq:zerocurv} 
can, once again, be analysed using the addition formula \eqref{eq:addform} for the Lam\'e function $\Phi_\kp$ and analyzed entry-by-entry. 
Applying this to each entry of both the left-hand side and right-hand side of \eqref{eq:zerocurv} we observe that
in all four entries a common factor containing the spectral parameter $\kp$ will drop out and that we are left with
the following four relations:
\bse\label{eq:compats0}\begin{eqnarray}
&& \wh{\ld}\mu \left[ \zeta(\wh{\wt{\xi}}-\wh{\xi}-\alpha)+\zeta(\wh{\xi}-\xi-\bb)
-\zeta(\wh{\wt{\xi}}+\wh{\xi}-\alpha)+\zeta(\wh{\xi}+\xi+\bb)\right] \nn \\
&& \qquad = \wt{\mu}\ld \left[ \zeta(\wh{\wt{\xi}}-\wt{\xi}-\bb)+\zeta(\wt{\xi}-\xi-\alpha)
-\zeta(\wh{\wt{\xi}}+\wt{\xi}-\bb)+\zeta(\wt{\xi}+\xi+\alpha)\right] \label{eq:compats0a}\\
 && \wh{\ld}\mu \left[ \zeta(\wh{\wt{\xi}}-\wh{\xi}-\alpha)+\zeta(\wh{\xi}+\xi-\bb)
-\zeta(\wh{\wt{\xi}}+\wh{\xi}-\alpha)+\zeta(\wh{\xi}-\xi+\bb)\right] \nn \\
&& \qquad = \wt{\mu}\ld \left[ \zeta(\wh{\wt{\xi}}-\wt{\xi}-\bb)+\zeta(\wt{\xi}+\xi-\alpha)
-\zeta(\wh{\wt{\xi}}+\wt{\xi}-\bb)+\zeta(\wt{\xi}+\xi-\alpha)\right] \label{eq:compats0b}\\
&& \wh{\ld}\mu \left[ \zeta(-\wh{\wt{\xi}}-\wh{\xi}-\alpha)+\zeta(\wh{\xi}-\xi-\bb)
-\zeta(-\wh{\wt{\xi}}+\wh{\xi}-\alpha)+\zeta(\wh{\xi}+\xi+\bb)\right] \nn \\
&& \qquad = \wt{\mu}\ld \left[ \zeta(\wh{\wt{\xi}}-\wt{\xi}-\bb)+\zeta(\wt{\xi}-\xi-\alpha)
-\zeta(\wh{\wt{\xi}}+\wt{\xi}-\bb)+\zeta(\wt{\xi}+\xi+\alpha)\right] \label{eq:compats0c}\\
&& \wh{\ld}\mu \left[ \zeta(-\wh{\wt{\xi}}-\wh{\xi}-\alpha)+\zeta(\wh{\xi}+\xi-\bb)
-\zeta(-\wh{\wt{\xi}}+\wh{\xi}-\alpha)+\zeta(\wh{\xi}-\xi+\bb)\right] \nn \\
&& \qquad = \wt{\mu}\ld \left[ \zeta(-\wh{\wt{\xi}}-\wt{\xi}-\bb)+\zeta(\wt{\xi}+\xi-\alpha)
-\zeta(-\wh{\wt{\xi}}+\wt{\xi}-\bb)+\zeta(\wt{\xi}-\xi+\alpha)\right] \nn \\
\label{eq:compats0d}
\end{eqnarray}\ese
Using the identity \eqref{eq:addform} 
these four relations can be rewritten as:
\bse\label{eq:compats}\begin{eqnarray}
&& \wh{\ld}\mu \frac{\sg(2\wh{\xi})\,\sg(\wh{\wt{\xi}}+\xi+\bb-\alpha) }{\sg(\wh{\wt{\xi}}-\wh{\xi}-\alpha)\,
\sg(\wh{\wt{\xi}}+\wh{\xi}-\alpha)\,\sg(\wh{\xi}-\xi-\bb)\,\sg(\wh{\xi}+\xi+\bb)} \nn \\
&& \qquad =\wt{\mu}\ld \frac{\sg(2\wt{\xi})\,\sg(\wh{\wt{\xi}}+\xi+\alpha-\bb) }{\sg(\wh{\wt{\xi}}-\wt{\xi}-\bb)\,
\sg(\wh{\wt{\xi}}+\wt{\xi}-\bb)\,\sg(\wt{\xi}-\xi-\alpha)\,\sg(\wt{\xi}+\xi+\alpha)} \label{eq:compatsa}\\
&&\wh{\ld}\mu \frac{\sg(2\wh{\xi})\,\sg(\wh{\wt{\xi}}-\xi+\bb-\alpha) }{\sg(\wh{\wt{\xi}}-\wh{\xi}-\alpha)\,
\sg(\wh{\wt{\xi}}+\wh{\xi}-\alpha)\,\sg(\wh{\xi}-\xi+\bb)\,\sg(\wh{\xi}+\xi-\bb)} \nn \\
&& \qquad =\wt{\mu}\ld \frac{\sg(2\wt{\xi})\,\sg(\wh{\wt{\xi}}-\xi+\alpha-\bb) }{\sg(\wh{\wt{\xi}}-\wt{\xi}-\bb)\,
\sg(\wh{\wt{\xi}}+\wt{\xi}-\bb)\,\sg(\wt{\xi}-\xi+\alpha)\,\sg(\wt{\xi}+\xi-\alpha)} \label{eq:compatsb}\\
&&\wh{\ld}\mu \frac{\sg(2\wh{\xi})\,\sg(\wh{\wt{\xi}}-\xi-\bb+\alpha) }{\sg(\wh{\wt{\xi}}-\wh{\xi}+\alpha)\,
\sg(\wh{\wt{\xi}}+\wh{\xi}+\alpha)\,\sg(\wh{\xi}-\xi-\bb)\,\sg(\wh{\xi}+\xi+\bb)} \nn \\
&& \qquad =\wt{\mu}\ld \frac{\sg(2\wt{\xi})\,\sg(\wh{\wt{\xi}}-\xi-\alpha+\bb) }{\sg(\wh{\wt{\xi}}-\wt{\xi}+\bb)\,
\sg(\wh{\wt{\xi}}+\wt{\xi}+\bb)\,\sg(\wt{\xi}-\xi-\alpha)\,\sg(\wt{\xi}+\xi+\alpha)} \label{eq:compatsc}\\
&&\wh{\ld}\mu \frac{\sg(2\wh{\xi})\,\sg(\wh{\wt{\xi}}+\xi-\bb+\alpha) }{\sg(\wh{\wt{\xi}}-\wh{\xi}+\alpha)\,
\sg(\wh{\wt{\xi}}+\wh{\xi}+\alpha)\,\sg(\wh{\xi}-\xi+\bb)\,\sg(\wh{\xi}+\xi-\bb)} \nn \\
&& \qquad =\wt{\mu}\ld \frac{\sg(2\wt{\xi})\,\sg(\wh{\wt{\xi}}+\xi-\alpha+\bb) }{\sg(\wh{\wt{\xi}}-\wt{\xi}+\bb)\,
\sg(\wh{\wt{\xi}}+\wt{\xi}+\bb)\,\sg(\wt{\xi}-\xi+\alpha)\,\sg(\wt{\xi}+\xi-\alpha)} .  \label{eq:compatsd}
\end{eqnarray}\ese
Eliminating $\ld$ and $\mu$, simply by dividing pairwise the relations over each other, we obtain directly the 
3-leg formulae. In fact, we obtain two seemingly different-looking equations for $\xi$, namely:
\bse\begin{equation}
\label{eq:3-leg}
\frac{\sg(\wt{\xi}-\xi+\alpha)\,\sg(\wt{\xi}+\xi-\alpha)}
{\sg(\wt{\xi}-\xi-\alpha)\,\sg(\wt{\xi}+\xi+\alpha)}\,
\frac{\sg(\wh{\xi}-\xi-\bb)\,\sg(\wh{\xi}+\xi+\bb)}
{\sg(\wh{\xi}-\xi+\bb)\,\sg(\wh{\xi}+\xi-\bb)}
=\frac{\sg(\wh{\wt{\xi}}-\xi-\gm)\,\sg(\wh{\wt{\xi}}+\xi+\gm)}
{\sg(\wh{\wt{\xi}}-\xi+\gm)\,\sg(\wh{\wt{\xi}}+\xi-\gm)}\ , 
\end{equation}
in which as before $\gm=\beta-\alpha$, and 
\begin{equation}
\label{eq:alt3-leg}
\frac{\sg(\wh{\wt{\xi}}-\wh{\xi}+\alpha)\,\sg(\wh{\wt{\xi}}+\wh{\xi}+\alpha)}
{\sg(\wh{\wt{\xi}}-\wh{\xi}-\alpha)\,\sg(\wh{\wt{\xi}}+\wh{\xi}-\alpha)}\,
\frac{\sg(\wh{\wt{\xi}}-\wt{\xi}-\bb)\,\sg(\wh{\wt{\xi}}+\wt{\xi}-\bb)}
{\sg(\wh{\wt{\xi}}-\wt{\xi}+\bb)\,\sg(\wh{\wt{\xi}}+\wt{\xi}+\bb)}
=\frac{\sg(\wh{\wt{\xi}}-\xi-\gm)\,\sg(\wh{\wt{\xi}}+\xi-\gm)}
{\sg(\wh{\wt{\xi}}-\xi+\gm)\,\sg(\wh{\wt{\xi}}+\xi+\gm)}\  ,
\end{equation} \ese
but actually these two equations are equivalent. 
The first equation \eqref{eq:3-leg} is identical to \eqref{eq:3leg}, namely the 3-leg form of the Adler lattice
equation.The second equation \eqref{eq:alt3-leg} is obtained from the first by interchanging 
$\xi\leftrightarrow\wh{\wt{\xi}}$, $\alpha\leftrightarrow\bb$, which is a symmetry of the equation. The equivalence 
between these two forms is made manifest by passing to the rational form \eqref{eq:ellform} of the equation, and 
the latter connection can be seen to be a consequence of an interesting identity given in the following statement.

\begin{prop}
For arbitrary (complex) variables X, Y, and Z, we have the following identity
\begin{eqnarray}\label{3-legeq}
\lefteqn{(X-\wp(\xi+\alpha))(Y-\wp(\xi-\beta))(Z-\wp(\xi-\alpha+\beta))\nn}\\
& &-t^2(X-\wp(\xi-\alpha))(Y-\wp(\xi+\beta))(Z-\wp(\xi+\alpha-\beta)) \nonumber \\
& &=s\big[(-a B-b A)(\wp(\xi)(XY+YZ+XZ)+XYZ)+(b^2 A+a^2 B)(Z\wp(\xi)+XY)\nonumber \\
& &+((b^2 A+a^2 B)-B(a-b)(a-c))(X\wp(\xi)+YZ)+(-A (b-a)(b-c)+(b^2 A+a^2 B))\nonumber \\
& &\times(Y\wp(\xi)+XZ)+(a B(a-b)(a-c)+bA(b-a)(b-c)-A b^3-B a^3)(\wp(\xi)+X+Y+Z)\nonumber \\
& &+A (b^2-(a-b)(c-b)))^2+B(a^2-(b-a)(c-a))^2-ABC(a-b)+(A+B)XYZ\wp(\xi)\big],\nn \\
\end{eqnarray}
in which
\begin{eqnarray}\label{eq:st}
t=\frac{\sigma(\xi-\alpha)\sigma(\xi+\beta)\sigma(\xi+\alpha-\beta)}{\sigma(\xi+\alpha)\sigma(\xi-\beta)\sigma(\xi-\alpha+\beta)},\quad s = \frac{1-t^2}{(A+B)\wp(\xi)-A b-a B}.
\end{eqnarray}
\end{prop}

\noindent 
A (computational) proof of the Proposition 3.1 is given in Appendix \ref{appendix:a}. 
Identifying $u=\wp(\xi)$, $X=\wt u=\wp(\wt{\xi})$, $Y=\wh u=\wp(\wh{\xi})$ and $Z=\wh{\wt u}=\wp(\widehat{\widetilde{\xi}})$, and using
\begin{equation}\label{eq:addforms2}
\wp(\xi)-\wp(\eta)=\frac{\sg(\eta+\xi)\,\sg(\eta-\xi)}{\sg^2(\eta)
\sg^2(\xi)}~~,
\end{equation}
it is not hard to see that the elliptic identity \eqref{3-legeq} relates the rational form of Adler's equation in the Weierstrass case \eqref{eq:ellform} and 3-leg \eqref{eq:3-leg}.
Since the Adler system \eqref{eq:ellform} is manifestly invariant under the replacements
$u\leftrightarrow\wh{\wt{u}}$, $\alpha\leftrightarrow\bb$ -- whilst \textit{not} interchanging
$\wt{u}$ and $\wh{u}$ -- (this being a particular aspect of the $D_4$-symmetry of the equation), the 
3-leg form \eqref{eq:3-leg} is also invariant under the parallel exchange on the level of the uniformising 
variables: $\xi\leftrightarrow\wh{\wt{\xi}}$, $\alpha\leftrightarrow\bb$.
This is the symmetry that connects the two forms \eqref{eq:3-leg} and \eqref{eq:alt3-leg}, which are
hence equivalent. 

\paragraph{Remark 1:}
The coefficients $\ld$ and $\mu$ are determined by the condition that the dynamical equation for the determinants of the Lax matrices
$L_\kp$, $M_\kp$ need to be trivially satisfied. Thus a possible choice for $\ld$ and $\mu$ is
to determine these factors such that ~$\det(L_\kp)$~ and ~$\det(M_\kp)$ are proportional to constants (i.e.
independent of $\xi$), which leads to the following expressions 
\begin{equation}\label{eq:muld}
\ld=\left(\frac{H(u,\wt{u},a)}{AU\wt{U}}\right)^{1/2}
\quad,\quad
\mu=\left(\frac{H(u,\wh{u},b)}{BU\wh{U}}\right)^{1/2}\  ,
\end{equation}
where $u=\wp(\xi)~,~U=r(u)=\wp'(\xi)$, and similary $\wt{u}=\wp(\wt{\xi})~,~\wt{U}=r(\wt{u})=\wp'(\wt{\xi})$, and
$\wh{u}=\wp(\wh{\xi})~,~\wh{U}=r(\wh{u})=\wp'(\wh{\xi})$.
The symmetric triquadratic function $H$ is given by
\begin{equation}
H(u,v,a)\equiv \left( uv+au+av+\frac{g_2}{4}\right)^2-(4a u v-g_3)(u+v+a)\   , \label{eq:H}
\end{equation}
and which can be obtained in the following form in terms of $\sigma$-function 
\begin{eqnarray}\label{eq:Hrel}
H(u,v,a)&=&(u-v)^2 \left[\frac{1}{4}\left(\frac{U-V}{u-v}\right)^2-(u+v+a)\right]
\left[\frac{1}{4}\left(\frac{U+V}{u-v}\right)^2-(u+v+a)\right] \nn \\ 
&=& \frac{\sg(\xi+\eta+\alpha)\,\sg(\xi+\eta-\alpha)\,\sg(\xi-\eta+\alpha)\,\sg(\xi-\eta-\alpha)}
{\sg^4(\xi)\,\sg^4(\eta)\,\sg^4(\alpha)}\  ,
\end{eqnarray}
in which $U^2\equiv r(u)$, $V^2\equiv r(v)$. We also have the expression in terms of the polynomial of the curve:  
\begin{equation}\label{eq:rr}
\left[ r(u)+r(a)-4(u-a)^2(u+v+a)\right]^2 -4 r(u)\,r(a)=16 (u-a)^2 H(u,v,a)\  .
\end{equation}
We further note at this point that the discriminant of the triquadratic in each argument factorises:
\begin{equation}\label{eq:Hdiscr}
H_v^2-2H\,H_{vv}=r(a)r(u)\  .
\end{equation}
In \cite{ABS2} the discriminant properties of affine-linear quadrilaterals and their relation with the 
corresponding biquadratics and their discriminants, were exploited to tighten the classification result of \cite{ABS}.

\paragraph{Remark 2:}
An alternative derivation of the $N=2$ case can be given by using the system of equations \eqref{eq:zetasystem}. In this case the variables $H_l$ and $K_l$ take on the following forms, setting $\xi_1=-\xi_2=\xi$:
\bse\label{eq:AdlHK}\begin{eqnarray}
H_1&=& \frac{\sg(\wh{\xi}-\wh{\wt{\xi}}+\alpha)\,\sg(\wh{\xi}+\wh{\wt{\xi}}+\alpha)
\,\sg(\wh{\xi}-\xi-\bb)\,\sg(\wh{\xi}+\xi-\bb)}{\sg(\wh{\xi}-\wt{\xi}-\gm)\,
\sg(\wh{\xi}+\wt{\xi}-\gm)\,\sg(2\wh{\xi})}\  , \label{eq:H1} \\
H_2&=& \frac{\sg(-\wh{\xi}-\wh{\wt{\xi}}+\alpha)\,\sg(-\wh{\xi}+\wh{\wt{\xi}}+\alpha)
\,\sg(-\wh{\xi}-\xi-\bb)\,\sg(-\wh{\xi}+\xi-\bb)}{\sg(-\wh{\xi}-\wt{\xi}-\gm)\,
\sg(-\wh{\xi}+\wt{\xi}-\gm)\,\sg(-2\wh{\xi})}\  , \label{eq:H2} \\
K_1&=& \frac{\sg(\wt{\xi}-\wh{\wt{\xi}}+\bb)\,\sg(\wt{\xi}+\wh{\wt{\xi}}+\bb)
\,\sg(\wt{\xi}-\xi-\alpha)\,\sg(\wt{\xi}+\xi-\alpha)}{\sg(\wt{\xi}-\wh{\xi}+\gm)\,
\sg(\wt{\xi}+\wh{\xi}+\gm)\,\sg(2\wt{\xi})}\  , \label{eq:K1} \\
K_2&=& \frac{\sg(-\wt{\xi}-\wh{\wt{\xi}}+\bb)\,\sg(-\wt{\xi}+\wh{\wt{\xi}}+\bb)
\,\sg(-\wt{\xi}-\xi-\alpha)\,\sg(-\wt{\xi}+\xi-\alpha)}{\sg(-\wt{\xi}-\wh{\xi}+\gm)\,
\sg(-\wt{\xi}+\wh{\xi}+\gm)\,\sg(-2\wt{\xi})}\  , \label{eq:K2}
\end{eqnarray}\ese
The identity $H_1+H_2=0$ upon inserting the above expressions yield the equation:
\be\label{eq:Hellform}
\left[ \frac{\sg(\wt{\xi}+\xi+\alpha)\,\sg(\wt{\xi}-\xi-\alpha)}
{\sg(\wt{\xi}+\xi-\alpha)\,\sg(\wt{\xi}-\xi+\alpha)}\right]^{\wh{\phantom{a}}}
\frac{\sg(\wh{\xi}+\xi-\bb)\,\sg(\wh{\xi}-\xi-\bb)}
{\sg(\wh{\xi}+\xi+\bb)\,\sg(\wh{\xi}-\xi+\bb)}=
\frac{\sg(\wt{\xi}+\wh{\xi}-\gm)\,\sg(\wt{\xi}-\wh{\xi}+\gm)}
{\sg(\wt{\xi}-\wh{\xi}-\gm)\,\sg(\wt{\xi}+\wh{\xi}+\gm)}\   ,
\ee
which is equivalent to the elliptic lattice system \eqref{eq:ellform} under the same
changes of variables as discussed before. In fact, \eqref{eq:Hellform} can be obtained
from \eqref{eq:3-leg} by interchanging: ~$\xi\leftrightarrow \wh{\xi}$~ and ~$\wh{\wt{\xi}}
\leftrightarrow \wt{\xi}$~. Similarly, the identity $K_1+K_2=0$ upon inserting the expressions
\eqref{eq:K1} and \eqref{eq:K2} for $K_1$ and $K_2$ yields a similar equation to \eqref{eq:Hellform} 
which can be obtained from \eqref{eq:3-leg} by interchanging: ~$\xi\leftrightarrow \wt{\xi}$~ and ~$\wh{\wt{\xi}}
\leftrightarrow \wh{\xi}$~. Thus, we recover from the scheme proposed in the previous
section the Adler system in the various 3-leg forms based at different vertices of the elementary quadrilateral.

\subsection{Case N=3:}

To generalise the results in the previous subsection to the rank 3 case, we consider the  
following form of a Lax representation on the lattice: 
\bse\label{eq:Ell3Lax}\begin{eqnarray}
&&\wt{\chi} =\left( \begin{array}{ccc}
h_1\Phi_{3\kp}(\wt{\xi}_1-\xi_1-\alpha)&h_2\Phi_{3\kp}(\wt{\xi}_1-\xi_2-\alpha) &
h_3\Phi_{3\kp}(\wt{\xi}_1-\xi_3-\alpha)\\
h_1\Phi_{3\kp}(\wt{\xi}_2-\xi_1-\alpha)&h_2\Phi_{3\kp}(\wt{\xi}_2-\xi_2-\alpha) &
h_3\Phi_{3\kp}(\wt{\xi}_2-\xi_3-\alpha) \\
h_1\Phi_{3\kp}(\wt{\xi}_3-\xi_1-\alpha)&h_2\Phi_{3\kp}(\wt{\xi}_3-\xi_2-\alpha) &
h_3\Phi_{3\kp}(\wt{\xi}_3-\xi_3-\alpha) \end{array} \right)\chi\   , \nn \\
\\
&&\wh{\chi}=\left( \begin{array}{ccc}
k_1\Phi_{3\kp}(\wh{\xi}_1-\xi_1-\bb)&k_2\Phi_{3\kp}(\wh{\xi}_1-\xi_2-\bb) &
k_3\Phi_{3\kp}(\wh{\xi}_1-\xi_3-\bb) \\
k_1\Phi_{3\kp}(\wh{\xi}_2-\xi_1-\bb)&k_2\Phi_{3\kp}(\wh{\xi}_2-\xi_2-\bb) &
k_3\Phi_{3\kp}(\wh{\xi}_2-\xi_3-\bb) \\
k_1\Phi_{3\kp}(\wh{\xi}_3-\xi_1-\bb)&k_2\Phi_{3\kp}(\wh{\xi}_3-\xi_2-\bb) &
k_3\Phi_{3\kp}(\wh{\xi}_3-\xi_3-\bb)
\end{array} \right)\chi\   , \nn \\
\end{eqnarray} \ese
subject to ~$\sum_{i=1}^3 h_i=\sum_{i=1}^3 k_i=0$~, and where the coefficients $h_j$,\;$k_j$ are some functions of the variables
$\xi_j$, and of their shifts. The compatibility conditions \eqref{eq:zerocurv} of this Lax pair results in a coupled set of Lax equations in terms of 
the three variables $\xi_j$ as we shall demonstrate by performing a similar type of analysis as in the case $N=2$, which in this case 
is understandably more involved. 

Eliminating\footnote{Equivalently, we could have eliminated $h_1$ or $h_2$ and $k_1$ or $k_2$
yielding equivalent results.}, $h_3=-h_1-h_2$ and $k_3=-k_1-k_2$  we obtain from \eqref{eq:sepeqsb} the following system of equations:
\begin{eqnarray}
&&\sum_{l=1}^2
\widehat{h}_l k_j\left[\zeta(\widehat{\widetilde{\xi}}_i-\widehat{\xi}_l-\alpha)+\zeta(\widehat{\xi}_l-\xi_j-\beta)-
\zeta(\widehat{\widetilde{\xi}}_i-\widehat{\xi}_3-\alpha)-\zeta(\widehat{\xi}_3-\xi_j-\beta)\right] \nn \\ 
&& =\sum_{l=1}^2
\widetilde{k}_l h_j\left[\zeta(\widehat{\widetilde{\xi}}_i-\widetilde{\xi}_l-\beta)+\zeta(\widetilde{\xi}_l-\xi_j-\alpha)-
\zeta(\widehat{\widetilde{\xi}}_i-\widetilde{\xi}_3-\beta)-\zeta(\widetilde{\xi}_3-\xi_j-\alpha)\right] \nn \\ 
&& \forall\; i,j = 1,2,3.
\end{eqnarray}
and using the addition formula ($\ref{eq:addform}$) we next get: 
\begin{eqnarray}\label{eq:LaxeqN=3}
&&\sum\limits_{l=1}^2 \widehat{h}_l
k_j\frac{\sg(\widehat{\widetilde{\xi_i}}-\widehat{\xi_l}-\widehat{\xi_3}+\xi_j-\alpha+\beta)\sg(\widehat{\xi_l}-\widehat{\xi_3})}
{\sg(\widehat{\widetilde{\xi_i}}-\widehat{\xi_l}-\alpha)\sg(\widehat{\xi_l}-\xi_j-
\beta)\sg(\widehat{\widetilde{\xi_i}}-\widehat{\xi_3}-\alpha)\sg(\widehat{\xi_3}-\xi_j-\beta)}=	
 \nn\\
&&\quad \quad \quad=\sum\limits_{l=1}^2 \widetilde{k}_l
h_j\frac{\sg(\widehat{\widetilde{\xi_i}}-\widetilde{\xi_l}-\widetilde{\xi_3}+\xi_j+\alpha-\beta)\sg(\widetilde{\xi_l}-\widetilde{\xi_3})}
{\sg(\widehat{\widetilde{\xi_i}}-\widetilde{\xi_l}-\beta)\sg(\widetilde{\xi_l}-\xi_j-\alpha)
\sg(\widehat{\widetilde{\xi_i}}-\widetilde{\xi_3}-\beta)\sg(\widetilde{\xi_3}-\xi_j-\alpha)}\nn
\\
&&\forall\; i,j = 1,2,3.
\end{eqnarray}
To write \eqref{eq:LaxeqN=3} in a more concise way, we denote the coefficients on the l.h.s. and r.h.s. of the equation as
$A_{ilj}\equiv A_{ilj}(\widehat{\widetilde{\xi}},\widehat{\xi},\xi;\alpha,\beta)$ and
$B_{ilj}\equiv B_{ilj}(\widehat{\widetilde{\xi}},\widetilde{\xi},\xi;\alpha,\beta)$ respectively. Noting the common factors
$h_j/k_j$ ($j=1,2,3$) in these equations, we next derive the system of six equations
\begin{eqnarray}\label{HKK}
\frac{h_j}{k_j}= \frac{A_{11j} \widehat{h}_1+A_{12j} \widehat{h}_2}{B_{11j} \widetilde{k}_1+B_{12j} \widetilde{k}_2}=\frac{A_{21j} \widehat{h}_1+A_{22j} \widehat{h}_2}{B_{21j} \widetilde{k}_1+B_{22j} \widetilde{k}_2}=\frac{A_{31j} \widehat{h}_1+A_{32j} \widehat{h}_2}{B_{31j} \widetilde{k}_1+B_{32j} \widetilde{k}_2}\nn \\	
(j = 1,2,3)\  .
\end{eqnarray}
We can rewrite the resulting set of relation \eqref{HKK} as
\begin{eqnarray}\label{eq:structure}
&& (A_{11j} B_{21j}-A_{21j} B_{11j})\widehat{h}_1 \widetilde{k}_1+(A_{11j} B_{22j}-A_{21j} B_{12j})\widehat{h}_1 \widetilde{k}_2\nn \\	
&&\qquad \qquad \qquad  \qquad \qquad+(A_{12j} B_{21j}-A_{22j} B_{11j})\widehat{h}_2 \widetilde{k}_1+(A_{12j} B_{22j}-A_{22j} B_{12j})\widehat{h}_2\widetilde{k}_2=0\nn \\	
&& (A_{11j} B_{31j}-A_{31j} B_{11j})\widehat{h}_1 \widetilde{k}_1+(A_{11j} B_{32j}-A_{31j} B_{12j})\widehat{h}_1 \widetilde{k}_2\nn \\	
&&\qquad \qquad \qquad \qquad \qquad+(A_{12j} B_{31j}-A_{32j} B_{11j})\widehat{h}_2 \widetilde{k}_1+(A_{12j} B_{32j}-A_{32j} B_{12j})\widehat{h}_2 \widetilde{k}_2=0\nn \\	
&& (A_{21j} B_{31j}-A_{31j} B_{21j})\widehat{h}_1 \widetilde{k}_1+(A_{21j} B_{32j}-A_{31j} B_{22j})\widehat{h}_1 \widetilde{k}_2\nn \\	
&&\qquad \qquad \qquad  \qquad \qquad+(A_{22j} B_{31j}-A_{32j} B_{21j})\widehat{h}_2 \widetilde{k}_1+(A_{22j} B_{32j}-A_{32j} B_{22j})\widehat{h}_2 \widetilde{k}_2=0 \nn \\	
&&\qquad \qquad \qquad(j = 1,2,3)\ ,
\end{eqnarray}
where
\bse\label{eq:AB}\begin{eqnarray}
A_{ilj} &=& \frac{\sg(\wh{\wt{\xi}}_i-\wh{\xi}_l-\wh{\xi}_3+\xi_j-\alpha+\beta)\,\sg(\wh{\xi}_l-\wh{\xi}_3)}{
\sg(\wh{\wt{\xi}}_i-\wh{\xi}_l-\alpha)\,\sg(\wh{\xi}_l-\xi_j-\beta)\,\sg(\wh{\wt{\xi}}_i-\wh{\xi}_3-\alpha)\,
\sg(\wh{\xi}_3-\xi_j-\beta)}\  , \\
B_{ilj} &=& \frac{\sg(\wh{\wt{\xi}}_i-\wt{\xi}_l-\wt{\xi}_3+\xi_j+\alpha-\beta)\,\sg(\wt{\xi}_l-\wt{\xi}_3)}{
\sg(\wh{\wt{\xi}}_i-\wt{\xi}_l-\beta)\,\sg(\wt{\xi}_l-\xi_j-\alpha)\,\sg(\wh{\wt{\xi}}_i-\wt{\xi}_3-\beta)\,
\sg(\wt{\xi}_3-\xi_j-\alpha)}\  .
\end{eqnarray}\ese 

\noindent 
We observe that these homogeneous bilinear systems for the variables $\widehat{h}_1$, $\widetilde{k}_1$, $\widehat{h}_2$ and 
$\widetilde{k}_2$ can be resolved by using Cayley's 3-dimensional $2\times2\times2$-hyperdeterminant \cite{Cayley}. Let us recall the 
general statement (cf. also \cite{GKZ}):

\begin{definition}
The hyperdeterminant of $2\times 2\times 2$ hyper-matrix $A=(a_{ijk})$ $(i,j,k=0,1)$ is given by:  
\begin{multline}\label{eq:hyper}
\mathrm{Det}(A)= \Big[\text{\rm det}
\begin{pmatrix}
a_{000}& a_{001}\\
a_{110}& a_{111}
\end{pmatrix}
+\text{\rm det}\begin{pmatrix}
a_{100} & a_{010}\\
a_{101} & a_{011}
\end{pmatrix}\Big]^2\\
-4\; \text{\rm det}\begin{pmatrix}
a_{000} & a_{001}\\
a_{010} & a_{011}
\end{pmatrix}
\text{\rm det}\begin{pmatrix}
a_{100}& a_{101}\\
a_{110}& a_{111}
\end{pmatrix} . 
\end{multline}
\end{definition} 

\noindent 
Its main property is the following: 
\begin{prop}\label{prop:homsyst} 
The hyper-determinant \eqref{eq:hyper} vanishes identically iff the following set of bilinear equations with six unknowns
\begin{eqnarray}\label{system}
&& a_{001}x_0y_0+a_{011}x_0y_1+a_{101}x_1y_0+a_{111}x_1y_1 = 0,\nn \\
&& a_{000}x_0y_0+a_{010}x_0y_1+a_{100}x_1y_0+a_{110}x_1y_1 = 0,\nn \\
&& a_{010}x_0z_0+a_{011}x_0z_1+a_{110}x_1z_0+a_{111}x_1z_1 = 0,\nn \\
&& a_{000}x_0z_0+a_{001}x_0z_1+a_{100}x_1z_0+a_{101}x_1z_1 = 0,\nn \\
&& a_{100}y_0z_0+a_{101}y_0z_1+a_{110}y_1z_0+a_{111}y_1z_1 = 0,\nn \\
&& a_{000}y_0z_0+a_{001}x_0z_1+a_{010}y_1z_0+a_{011}y_1z_1 = 0,
\end{eqnarray}
has a non-trivial solution (i.e., for which none of the vectors $\boldsymbol{x}=(x_0,x_1)$, $\boldsymbol{y}=(y_0,y_1)$, $\boldsymbol{z}=(z_0,z_1)$ are equal to the zero vector).  
\end{prop}
A proof of this statement can be found in \cite{ShadHassibi}. The cubic hyper-matrix $A$ can be illustrated by the following diagram of entries 
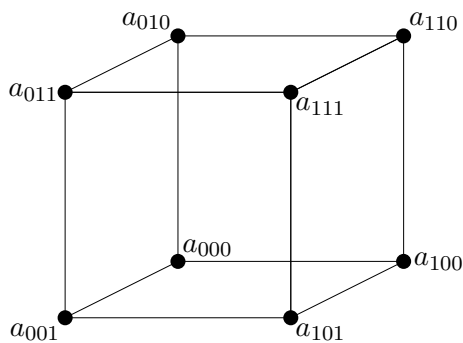
\begin{figure}[ht]
\centering
\begin{tikzpicture}[every node/.style={minimum size=0.5cm},on grid]
 \draw[black, very thin] (1.5,0.75) -- (4.5,0.75);
 \draw[black, very thin] (1.5,0.75) -- (1.5,3.75);
\begin{scope}[every node/.append style={yslant=0.5},yslant=0.5]
 \draw[black , very thin] (0,0) -- (1.5,0);
\end{scope}
 \draw [black, very thin](0,0) rectangle (3,3);
\begin{scope}[every node/.append style={yslant=0.5,xslant=-1},xslant=2]
 \draw [black, very thin](-6,3) rectangle (-3,3.75);
\end{scope}
\begin{scope}[every node/.append style={yslant=0.5},yslant=0.5]
 \draw[black, very thin] (3,-1.5) rectangle (4.5,1.5);
\end{scope}
 \fill (4.5,3.75) circle (0.1) node[above right=-0.09] {$a_{110}$};
 \fill[black, very thin] (1.5,0.75) circle (0.1) node[above right=-0.1] {$a_{000}$};
 \fill(0,3) circle (0.1) node[left=-0.05] {$a_{011}$};
 \fill (1.5,3.75) circle (0.1) node[above left=-0.08] {$a_{010}$};
 \fill(3,0) circle (0.1) node[below right=-0.1] {$a_{101}$};
 \fill[black] (0,0) circle (0.1) node[below left=-0.1] {$a_{001}$};
 \fill[black] (4.5,0.75) circle (0.1) node[right] {$a_{100}$};
 \fill[black] (3,3) circle (0.1) node[below right=-0.10] {$a_{111}$};
\end{tikzpicture}
\caption{Cayley Cube}
	\label{Cube}
 \end{figure}

In the case at hand, the components $a_{ijk}$ can be readily identified by comparing \eqref{eq:structure} with the system
\eqref{system} and the variables $x_i$, $y_j$ with the $\widehat{h}_i$ and $\widetilde{k}_j$ respectively.
Noting that these particular coefficients are all $2\times 2$ determinants, it turns out that the following \textit{compound
theorem for hyper-determinants} is directly applicable.


\begin{lemma}[Compound Theorem for $2\times2\times2$ hyper-determinants] The following identity holds for the compound hyper-
determinants of format $2\times 2\times 2$:
\begin{eqnarray}
 \Big(\;\left| \begin{array}{cc} \left| \begin{array}{cc} a& a'' \\
 b& b'' \\
 \end{array}\right|&\left| \begin{array}{cc} a'& a'' \\
 d'& d'' \\
\end{array}\right|\\
\noalign{\vskip12pt}
\left| \begin{array}{cc} c& c'' \\
 b& b'' \\
\end{array}\right|&\left| \begin{array}{cc} c'& c'' \\
 d'& d'' \\
\end{array}\right| \end{array}\right|+\left| \begin{array}{cc} \left| \begin{array}{cc} a'& a'' \\
 b'& b'' \\
 \end{array}\right|&\left| \begin{array}{cc} a& a'' \\
 d& d'' \\
\end{array}\right|\\
\noalign{\vskip12pt}
\left| \begin{array}{cc} c'& c'' \\
 b'& b'' \\
\end{array}\right|&\left| \begin{array}{cc} c& c''\nn \\
 d& d'' \\
\end{array}\right| \end{array}\right|\;\Big)^2 &&\\
\noalign{\vskip12pt}
-4\left| \begin{array}{cc} \left| \begin{array}{cc} a& a'' \\
 b& b'' \\
 \end{array}\right|&\left| \begin{array}{cc} a& a'' \\
 d& d'' \\
\end{array}\right|\\
\noalign{\vskip12pt}
\left| \begin{array}{cc} c& c'' \\
 b& b'' \\
\end{array}\right|&\left| \begin{array}{cc} c& c'' \\
 d& d'' \\
\end{array}\right| \end{array}\right|.\left| \begin{array}{cc} \left| \begin{array}{cc} a'& a'' \\
 b'& b'' \\
 \end{array}\right|&\left| \begin{array}{cc} a'& a'' \\
 d'& d'' \\
\end{array}\right|\\
\noalign{\vskip12pt}
\left| \begin{array}{cc} c'& c'' \\
 b'& b'' \\
\end{array}\right|&\left| \begin{array}{cc} c'& c'' \\
 d'& d'' \\
\end{array}\right| \end{array}\right|
&=&\left| \begin{array}{cc} \left| \begin{array}{cc} a& a'' \\
 c& c'' \\
 \end{array}\right|&\left| \begin{array}{cc} b& b'' \\
 d& d'' \\
\end{array}\right|\nn \\
\noalign{\vskip12pt}
\left| \begin{array}{cc} a'& a'' \\
 c'& c'' \\
\end{array}\right|&\left| \begin{array}{cc} b'& b'' \\
 d'& d'' \\
\end{array}\right| \end{array}\right|^2. \nn \\
\noalign{\vskip12pt}
\label{CTheo}
\end{eqnarray}
\end{lemma}
\normalsize
\begin{proof}
This can be established by direct computation. Assuming w.l.o.g. that the entries $a''$, $b''$,$c''$.$d''$ are all nonzero,
we can take out the common product $(a''b''c''d'')^2$ from all terms on the left-hand side.
Denoting all the ratios $a/a''$, $a'/a''$ by capitals $A$, $A'$ etc., and noting that the $2\times 2$ determinant
$\left| \begin{array}{cc}  a/a''& 1 \\ b/b''&1 \\ \end{array}\right|$ is simply given by $A-B$ (and in a similar way the other determinants occurring in the expression on the left-hand side), then the left-hand side of \eqref{CTheo} is representable by
\small
\begin{eqnarray*}
 && a''^2\;b''^2\;c''^2\;d''^2\Bigg[\Bigg( \left| \begin{array}{cc} A-B& A'-D' \\
 C-B& C'-D' \\
 \end{array}\right|+\left| \begin{array}{cc} A'-B'& A-D \\
 C'-B'& C-D \\
 \end{array}\right|\Bigg)^2 \nn \\
&&\qquad \qquad \qquad\qquad-4 \left| \begin{array}{cc} A-B& A-D \\
 C-B& C-D \\
 \end{array}\right|. \left| \begin{array}{cc} A'-B'& A'-D' \\
 C'-B'& C'-D' \\
 \end{array}\right|\Bigg]\  .
\end{eqnarray*}
\normalsize
Computing the expression between brackets, we observe that it can be simplified to:
\begin{eqnarray*}
&& \left( (A-C)(B'-D')+(D-B)(C'-A')\right)^2 -4(A-C)(B-D)(A'-C')(B'-D') \\
&& =\left|\begin{array}{cc} A-C & B-D \\ A'-C' & B'-D'\end{array}\right|^2\  ,
\end{eqnarray*}
which leads to the desired result.
\end{proof}

This compound theorem to the best of our knowledge is a new result in the theory of hyper-determinants. It seems intimately linked 
to the structure of the linear equations (the Lax relations) from which it originate in the present context, and there may be 
analogues for the case of higher rank hyper-determinants (this is currently under investigation). A connection between 
hyper-determinants and minors of symmetric matrices was established in \cite{Holtz}, but it is not clear whether (and if so how) 
those results are related to the above proposition.  

Identifying the coefficients of the system of homogeneous equations \eqref{eq:structure} as entries of a $2\times 2\times 2$ 
hyper-determinant, we observe that the structure of this hyper-determinant is exactly of the form as given in Lemma 3.1, and 
hence we have the following immediate corollary.  

\begin{prop}\label{prop:redsyst} 
Identifying the 8 entries $(a_{ijk})_{i,j,k=0,1}$ by comparing the first two equations of \eqref{system} with the system of 
equations \eqref{eq:structure}, the hyper-determinant takes the form as given by the compound 
theorem Lemma 3.1, and hence reduces to a perfect square of the form:
\begin{eqnarray}\label{eq:N=3systems}
&& \left| \begin{array}{cc} 
\begin{vmatrix}
A_{ilj}&A_{il'j} \\
A_{i''lj}&A_{i''l'j}
\end{vmatrix} & 
\begin{vmatrix}
A_{i'lj}&A_{i'l'j} \\
A_{i''lj}&A_{i''l'j}
\end{vmatrix}\\ 
 & \\ 
\begin{vmatrix}
B_{ilj}& B_{il'j}\\
B_{i''lj}& B_{i''l'j}
\end{vmatrix}
& 
\begin{vmatrix}
 B_{i'lj}&B_{i'l'j}\\
B_{i''lj}&B_{i''l'j}
\end{vmatrix}    
\end{array}\right|^2\qquad  
 (j = 1,2,3),
\end{eqnarray}
where
\begin{eqnarray}\label{det:A}
\begin{vmatrix}
A_{ilj}&A_{il'j}\\
A_{i''l j}&A_{i''l'j}
\end{vmatrix}
&=&\frac{\sg(\wh{\xi}_l-\wh{\xi}_{3})\,\sg(\wh{\xi}_{l'}-\wh{\xi}_{3})\,\sg(\wh{\xi}_l-\wh{\xi}_{l'})}{\sg(\wh{\wt{\xi}}_i-\wh{\xi}_l-\alpha)\,\sg(\wh{\wt{\xi}}_i-\wh{\xi}_{l'}-\alpha)\,
\sg(\wh{\wt{\xi}}_{i''}-\wh{\xi}_l-\alpha)\sg(\wh{\wt{\xi}}_{i''}-\wh{\xi}_{l'}-\alpha)}\nn \\
&\times&\frac{\sg(\wh{\wt{\xi}}_i-\wh{\wt{\xi}}_{i''})\,
\sg(\wh{\wt{\xi}}_i+\wh{\wt{\xi}}_{i''}-\wh{\xi}_l-\wh{\xi}_{l'}-\wh{\xi}_{3}+\xi_j-2\alpha+\beta)}
{\sg(\wh{\wt{\xi}}_i-\wh{\xi}_{3}-\alpha)\,\sg(\wh{\wt{\xi}}_{i''}-\wh{\xi}_{3}-\alpha)\,
\sg(\wh{\xi}_l-\xi_j-\beta)\,\sg(\wh{\xi}_{l'}-\xi_j-\beta)\,\sg(\wh{\xi}_{3}-\xi_j-\beta)},\nn \\
\end{eqnarray}
in which we can set $i,i'=1,2$, $l,l'=1,2\neq 3$, and where we naturally should take $i''=3$. 
A similar expression for the corresponding determinant in terms of the $B_{ilj}$ as given \eqref{det:A} 
interchanging $\alpha$ and $\beta$ and the shifts $\wt{\phantom{a}}$ and $\wh{\phantom{a}}$~. 
\end{prop} 

\noindent 
The form \eqref{det:A} of the relevant $2\times 2$ determinants, using the expressions for the entries \eqref{eq:AB}, is 
computed in Appendix \ref{appendix:c}. 

We apply now the compound theorem Lemma 3.1  to the system of homogeneous equations \eqref{eq:structure}. In fact from that system 
of equations it follows that the ratios $\wh{h}_i/\wh{h}_j$ and $\wt{k}_i/\wt{k}_j$ obey quadratic equations whose 
discriminant, by virtue of the compound theorem, is a perfect square. Thus, those ratios can be obtained in a rather 
simple form. We distinguish between the two cases: {\it i)} the hyper-determinant in question, i.e. 
the determinant \eqref{eq:N=3systems}, vanishes, and {\it ii)} the hyper-determinant is non-zero. 

\subsubsection*{\textit{i)} Case \eqref{eq:N=3systems}$=0$}

In this case the resulting set of equations is given by the vanishing of the hyper-determinant, i.e. the set of  
equations: 
\be\label{eq:detA=0} 
\begin{vmatrix}
A_{ilj}&A_{i l'j} \\
A_{i''lj}&A_{i''l'j}
\end{vmatrix} \,
\begin{vmatrix}
 B_{i'lj}&B_{i'l'j}\\
B_{i''lj}&B_{i''l'j}
\end{vmatrix}= 
\begin{vmatrix}
A_{i'lj}&A_{i'l'j} \\
A_{i''lj}&A_{i''l'j}
\end{vmatrix}\,  
\begin{vmatrix}
B_{ilj}& B_{il' j}\\
B_{i'' l j}& B_{i'' l'j}
\end{vmatrix}
\ee 
Inserting the explicit expression \eqref{det:A}, and its counterpart in terms of the quantities $B_{ilj}$, 
into \eqref{eq:detA=0} we obtain the relations 
\begin{eqnarray}\label{eq:AdlBSQ}
\frac{\sg(\wh{\wt{\xi}}_i+\wh{\wt{\xi}}_{i''}-\wh{\xi}_l-\wh{\xi}_{l'}-\wh{\xi}_3+\xi_j+\beta-2\alpha)}
{\sg(\wh{\wt{\xi}}_{i'}+\wh{\wt{\xi}}_{i''}-\wh{\xi}_l-\wh{\xi}_{l'}-\wh{\xi}_3+\xi_j+\beta-2\alpha)}\,
\frac{\sg(\wh{\wt{\xi}}_{i'}-\wh{\xi}_l-\alpha)\sg(\wh{\wt{\xi}}_{i'}-\wh{\xi}_{l'}-\alpha)\sg(\wh{\wt{\xi}}_{i'}-\wh{\xi}_3-\alpha)}
{\sg(\wh{\wt{\xi}}_i-\wh{\xi}_l-\alpha)\sg(\wh{\wt{\xi}}_i-\wh{\xi}_{l'}-\alpha)\sg(\wh{\wt{\xi}}_i-\wh{\xi}_3-\alpha)}
\qquad \qquad\nn \\ \nn \\
=\frac{\sg(\wh{\wt{\xi_{i}}}+\wh{\wt{\xi}}_{i''}-\wt{\xi}_l-\wt{\xi}_{l'}-\wt{\xi}_3+\xi_j+\alpha-2\beta)}
{\sg(\wh{\wt{\xi}}_{i'}+\wh{\wt{\xi}}_{i''}-\wt{\xi}_l-\wt{\xi}_{l'}-\wt{\xi}_3+\xi_j+\alpha-2\beta)}\,
\frac{\sg(\wh{\wt{\xi}}_{i'}-\wt{\xi}_l-\beta)\sg(\wh{\wt{\xi}}_{i'}-\wt{\xi}_{l'}-\beta)\sg(\wh{\wt{\xi}}_{i'}-\wt{\xi}_3-\beta)}
{\sg(\wh{\wt{\xi}}_i-\wt{\xi}_l-\beta)\sg(\wh{\wt{\xi}}_i-\wt{\xi}_{l'}-\beta)\sg(\wh{\wt{\xi}}_i-\wt{\xi}_3-\beta)} \qquad\nn \\	
\qquad \qquad\qquad \qquad\qquad \qquad \qquad (j = 1,2,3)\  ,  \qquad
\end{eqnarray}
where again we can set $i,i'=1,2$, $l,l'=1,2\neq 3$, and where we naturally should take $i''=3$. 
The set of relations \eqref{eq:AdlBSQ} is a coupled system of three quadrilateral equations (for $j=1,2,3$) of 3-leg type, i.e. 
in terms of three independent variables which reside in the arguments of the Weierstrass $\sg$-functions\footnote{An equivalent system 
of equations would have been obtained if, rather than eliminating $h_3$ and $k_3$ in its derivation, we would have eliminated one of the 
other variables among the coefficients $h_l$ and $k_l$.}. We note that all three equations (for $j=1,2,3$) have a common factor, which 
in the case of a further reduction $\xi_1+\xi_2+\xi_3=0$(mod period lattice) involves only the "long legs" (i.e. the differences over the diagonal). 
Thus, this system of equations may be too simple to figure as a proper candidate for a higher-rank analogue of the Adler lattice equation.

\subsubsection*{\textit{ii)} Case \eqref{eq:N=3systems}$\neq 0$} 

As a consequence of the compound theorem, Lemma 3.1, the hyper-determinant in the case at hand is a perfect square. Thus, going back 
to the system \eqref{eq:structure}, by first eliminating the ratio $\wh{h}_i/\wh{h}_j$, we obtain a quadratic for the ratio 
$\wt{k}_i/\wt{k}_j$, ($i,j=1,2$) from which the latter can be solved using the fact that the discriminant of the quadratic (which 
coincides with the hyper-determinant) is a perfect square. Thus, we get rather manageable expressions for the solutions 
of the mentioned ratios in terms of the $2\times 2$ determinants involving the expressions $A_{ilj}$ and $B_{ilj}$. The 
result of this computation is the following: 

\begin{prop}\label{prop:hksols} 
If the expression \eqref{eq:N=3systems} is non-vanishing, we have the following solutions of the system \eqref{eq:structure} 
given in terms of the ratios (i.e., up to a common multiplicative factor) 
\bse\label{eq:hkresol}\begin{eqnarray}
&{\rm either}&\qquad \frac{\wh{h}_1}{\wh{h}_2}=-\frac{A_{32j}}{A_{31j}}\quad {\rm together\ \ with}\quad  
\frac{\wt{k}_1}{\wt{k}_2}=-\frac{B_{32j}}{B_{31j}}\  , \label{eq:hkresola}\\ 
&{\rm or}&\quad 
\frac{\wh{h}_1}{\wh{h}_2}=
-\frac{\left|\begin{array}{ccc} B_{11j} & A_{12j}& B_{12j}\\ B_{21j} & A_{22j}& B_{22j}\\ B_{31j} & A_{32j}& B_{32j}\end{array}\right|}
{\left|\begin{array}{ccc} B_{11j} & A_{11j}& B_{12j}\\ B_{21j} & A_{21j}& B_{22j}\\ B_{31j} & A_{31j}& B_{32j}\end{array}\right|}
\quad {\rm together\ \ with}\quad 
\frac{\wt{k}_1}{\wt{k}_2}=
-\frac{\left|\begin{array}{ccc} A_{11j} & A_{12j}& B_{12j}\\ A_{21j} & A_{22j}& B_{22j}\\ A_{31j} & A_{32j}& B_{32j}\end{array}\right|}
{\left|\begin{array}{ccc} A_{11j} & A_{12j}& B_{11j}\\ A_{21j} & A_{22j}& B_{21j}\\ A_{31j} & A_{32j}& B_{31j}\end{array}\right|}\ . 
\nn \\ 
&& (j=1,2,3) \label{eq:hkresolb}
\end{eqnarray} \ese 
\end{prop} 

\noindent 
The proof, once again, is by direct computation and involves some determinantal manipulations.  


The system of equations resulting from \eqref{eq:hkresola}, inserting the explicit expressions for the quantities $A$ and 
$B$  from \eqref{eq:AB} reads as follows
\bse\begin{eqnarray}
\frac{\wh{h}_1}{\wh{h}_2}=-\frac{\sg(\wh{\wt{\xi}}_3-\wh{\xi}_2-\wh{\xi}_3+\xi_j-\alpha+\beta)\,\sg(\wh{\wt{\xi}}_3-\wh{\xi}_1-\alpha)\,\sg(\wh{\xi}_1-\xi_j-\beta)\,\sg(\wh{\xi}_2-\wh{\xi}_3)}{\sg(\wh{\wt{\xi}}_3-\wh{\xi}_1-\wh{\xi}_3+\xi_j-\alpha+\beta)\,\sg(\wh{\wt{\xi}}_3-\wh{\xi}_2-\alpha)\,\sg(\wh{\xi}_2-\xi_j-\beta)\,\sg(\wh{\xi}_1-\wh{\xi}_3)},
\end{eqnarray}\ese 
and
\bse\begin{eqnarray}
\frac{\wt{k}_1}{\wt{k}_2}=-\frac{\sg(\wh{\wt{\xi}}_3-\wt{\xi}_2-\wt{\xi}_3+\xi_j+\alpha-\beta)\,\sg(\wh{\wt{\xi}}_3-\wt{\xi}_1-\beta)\,\sg(\wt{\xi}_1-\xi_j-\alpha)\,\sg(\wt{\xi}_2-\wt{\xi}_3)}{\sg(\wh{\wt{\xi}}_3-\wt{\xi}_1-\wt{\xi}_3+\xi_j+\alpha-\beta)\,\sg(\wh{\wt{\xi}}_3-\wt{\xi}_2-\beta)\,\sg(\wt{\xi}_2-\xi_j-\alpha)\,\sg(\wt{\xi}_1-\wt{\xi}_3)}.\\
 (j=1,2,3)\nn
\end{eqnarray}\ese 
Inserting the expressions of \eqref{eq:AB} into the system of equations (comprising the equations for different values of 
$j=1,2,3$)  \eqref{eq:hkresolb} yields a more complicated system of quadrilateral elliptic 3-leg type of equations, which we 
have so far not been able to simplify\footnote{Note that the $3\times 3$ determinants in \eqref{eq:hkresolb} are almost, but 
not quite, of Frobenius (i.e., elliptic Cauchy) type.}. The problem of finding a rational form for the system of equations, as 
well as verifying their reducibility under the additional constraint $\xi_1+\xi_2+\xi_3=0$(mod period lattice) is currently under 
investigation.  We believe that the latter system of equations may correspond to the proper higher-rank analogue of Adler's lattice 
equation, but further work is needed to underpin that assertion.

\subsection*{Remark} 
In this paper we have proposed higher-rank lattice systems which by the construction we think of as natural analogues of Adler's lattice 
equation in 3-leg form. It is desirable to find their rational forms, similar to those of the Adler equation, i.e., 
either in the Weierstrass case given by \eqref{eq:ellform} or in the Jacobi case \eqref{Jacobi}, because in those forms  
the ${\rm D}_4$ symmetries of the equation are manifest. We mention here that the Jacobi form of Adler's equation \eqref{Jacobi}  
can be written in a remarkably succinct way using spin vectors. 

Introducing a ``spin matrix'' in the following way:    
\be  
\boldsymbol{G}\,\boldsymbol{\sigma}_3\,\boldsymbol{G}^{-1}=\boldsymbol{S}\cdot\boldsymbol{\sigma}\ , \quad {\rm where}\quad 
\boldsymbol{G}=\left(\begin{array}{cc} 1&1\\ u&v\end{array}\right)\  , 
\ee 
using the basis of the standard Pauli matrices $\boldsymbol{\sigma}=(\sg_1,\sg_2,\sg_3)$, we can identify 
a (normalised) spin vector as 
\be
\boldsymbol{S}(u,v)=\frac{1}{v-u}\left( uv-1,-i(uv+1),u+v\right)\  , \quad |\boldsymbol{S}|^2=
\boldsymbol{S}\cdot\boldsymbol{S}=1\  .   
\ee
Such a spin representation has been used in connection with the Landau-Lifschitz equations, cf. e.g. \cite{Adler3}. 
We have now the following statement: 

\begin{prop}
Adler's lattice equation in Jacobi form, i.e. \eqref{Jacobi}, can be represented in the following spin form:
\be\label{eq:Adlerspin} 
J_0+\boldsymbol{S}(v,\wt{v})\cdot\boldsymbol{J}\boldsymbol{S}(\wh{v},\wh{\wt{v}})=0\  , 
\ee 
in which the coefficient (anisotropy parameters) comprising $J_0$ and the $3\times 3$ diagonal matrix 
$\boldsymbol{J}={\rm diag}(J_1,J_2,J_3)$ are given by 
\be
J_0= \frac{q-r}{2}\ , \quad J_1=p\frac{1-qr}{2} \  ,\quad  J_2= p\frac{1+qr}{2}\  , \quad J_3=\frac{q+r}{2}\  , 
\ee 
with $r=(pQ-qP)/(1-p^2q^2)$. 

\end{prop}

The proof is by direct computation, writing out the components and identifying the various combinations of terms with 
the ones occurring in \eqref{Jacobi}. Obviously, the particular way \eqref{eq:Adlerspin} of writing the equation is 
not unique: it is subject to the ${\rm D}_4$ symmetries of the quadrilateral both in how the spin variables depend 
on the variables $v$ on the vertices and in how the anisotropy parameters depend on the lattice parameters. 

This observation suggests that the search for a rational form of higher-rank Adler lattice systems may involve higher spin variables, 
which are constructed in the following way. Using a basis of $GL_3$ given by the set of matrices\footnote{Following \cite{Bel} this can obviously be readily generalized to 
the case of $GL_N$.} $\{ \boldsymbol{I}_{n_1,n_2}\,|\,n_1,n_1\in\mathbb{Z}_3\}$, 
where ~$\boldsymbol{I}_{\boldsymbol{n}}=\boldsymbol{I}_{n_1,n_2}:=\boldsymbol{\Sigma}^{n_1}\boldsymbol{\Omega}^{n_2}=\oa^{n_1n_2}\boldsymbol{\Omega}^{n_2}\boldsymbol{\Sigma}^{n_1}$~  
are defined in terms of the elementary matrices    
\[ \boldsymbol{\Omega}=\left(\begin{array}{ccc}1 & & \\ 
& \omega &  \\ & & \omega^2 
\end{array}\right) 
\  , \quad 
\boldsymbol{\Sigma}=\left(\begin{array}{ccc}0 & 1 & 0 \\ 
0 & 0 & 1 \\ 1 & 0& 0   
\end{array}\right) 
\] 
and where $\omega$ is the 3$^{\rm d}$ root of unity, $\omega=\exp(2\pi i/3)$. 
These matrices obey the following relation 
\[ \boldsymbol{I}_{\boldsymbol{n}}\,\boldsymbol{I}_{\boldsymbol{m}}=\omega^{-n_2m_1}\boldsymbol{I}_{\boldsymbol{n}+\boldsymbol{m}}\  , \quad {\rm with}\quad 
\boldsymbol{n},\boldsymbol{m}\in\mathbb{Z}_3^2\  , \quad {\rm and} \quad \boldsymbol{I}_{\boldsymbol{n}}^\dagger=\oa^{-n_1n_2}\boldsymbol{I}_{-\boldsymbol{n}}\  . 
\] 
where the $\phantom{a}^\dagger$ means Hermitian conjugation. 
Thus, introducing the (traceless) spin matrix  
\[ 
\boldsymbol{S}:=\sum_{\boldsymbol{n}\in\mathbb{Z}_3^2\atop \boldsymbol{n}\neq(0,0)}\,S_{\boldsymbol{n}}\cdot\boldsymbol{I}_{\boldsymbol{n}}=\boldsymbol{G}\,\boldsymbol{\Omega}\,\boldsymbol{G}^{-1}\  , \quad {\rm with}\quad 
\boldsymbol{G}=\left(\begin{array}{ccc} 1&1&1\\ u_1&v_1&w_1 \\ u_2 &v_2 &w_2\end{array}\right)\  , 
\quad \boldsymbol{\Omega}=\left(\begin{array}{ccc} 1&&\\ &\omega& \\ &&\omega^2\end{array}\right)\  , \quad  
\] 
which is normalised through identity $\boldsymbol{S}^3=\boldsymbol{1}$, we can identify an 8-component spin vector $\boldsymbol{S}=(S_{n_1,n_2})$: 

in terms of the  
a 8-component vector $\boldsymbol{S}=(S_{n_1,n_2})$ where $n_1,n_2=0,1,2$, $(n_1,n_2)\ne (0,0)$, whose entries 
can be identified in the following way: 
\[ 
\boldsymbol{S}= \frac{1}{\boldsymbol{u}\cdot(\boldsymbol{v}\times\boldsymbol{w})}\, 
\left(\boldsymbol{u},\boldsymbol{v},\boldsymbol{w}\right)\boldsymbol{\Omega}\left( \begin{array}{c} 
(\boldsymbol{v}\times\boldsymbol{w})^T \\ (\boldsymbol{w}\times\boldsymbol{u})^T \\ (\boldsymbol{u}\times\boldsymbol{v})^T 
\end{array}\right)\  , \] 
from which the 8 spin components $S_{n_1,n_2}=S_{n_1,n_2}(\boldsymbol{u},\boldsymbol{v},\boldsymbol{w})$, 
($n_1,n_2\in\mathbb{Z}_3$, $(n_1,n_2)\neq(0,0)$), can be inferred comparing the entries 
of the matrices on the left-hand and right-hand sides. We aim at exploring the possibility of writing the higher-rank lattice systems 
in this representation.

\section{Discussion}

In this paper we have proposed and investigated a general class of higher-rank elliptic Lax representations for 
systems of partial difference equations on the 2D lattice. Distinguishing between what we called spin-zero (generalizations 
of Adler's lattice equation) and spin-nonzero (generalized landau-Lifschitz type) models, we gave the general 
structure of the resulting equations (from the compatibility conditions) for the latter, but concentrated mainly 
on the former case for $N=2$ and $N=3$.  For $N=2$ we have shown that the Lax systems leads indeed to Adler's lattice 
equation in its 3-leg form  (for the Weierstrass class) and we have analysed how these results generalize to the 
case $N=3$ (as a representative example for the higher-rank case). Having established the resulting systems of equations, 
generalizing Adler's 3-leg form, further work is needed to properly identify those systems. Thus, in further study we 
will investigate their rational and hyperbolic degenerations, as well as their continuum limits. A possible outcome would be 
to establish a connection with a differential system obtained by O. Mokhov in the 1980s, \cite{Mokh1}, arising from third
order commuting differential operators defining rank 3 vector bundles over an elliptic curve, cf. \cite{Mokh0}.

In our view, the significance of the results of this paper is not only to add a new class of elliptic type of integrable systems
to our already substantial zoo of such systems, but to depart from the rather restrictive confinement of $2\times 2$ systems 
to which all ABS type systems, \cite{ABS}, belong. To obtain good insights in the essential structures behind (discrete and continuous) 
integrable systems, such departures into the multi-component cases are necessary. In the present paper we concentrated mostly on
the spin-zero case, while the elaboration of the spin non-zero case is the subject of a future publication, some initial
results of which were already presented in \cite{SRY}. As a direction for the future, establishing connections, if any, 
with the recently found master-solutions of the quantum Yang-Baxter equations, \cite{BazhSerg}, may be of interest.

\section*{Acknowledgement}

ND is supported by the Turkish Ministry of National Education. FWN is thankful for the hospitality of the Department of Mathematics of
Shanghai University, during a visit where the current paper was finalized. He was partially supported by the
EPSRC responsive mode grants EP/I002294/1 and EP/I038683/1. SY is supported by Thailand Research Fund (TRF) under grand number: TRG5680081.

\appendix
\section{Proof of the Q4 3-leg identity}
\label{appendix:a}
\numberwithin{equation}{section}

The proof of the elliptic identity \eqref{3-legeq} can be achieved directly by showing that the coefficients of each monomials $1, X, Y, Z, XY, XZ, YZ$ and $XYZ$ of the
identity are equivalent. By expanding the left-hand side of the identity as
\begin{eqnarray}\label{LHS}
LHS:=\lefteqn{(1-t^2)X Y Z +(t^2\wp(\xi-\alpha)-\wp(\xi+\alpha))Y Z+(t^2\wp(\xi+\beta)-\wp(\xi-\beta))X Z\nn}\\
&+&(t^2\wp(\xi+\alpha-\beta)-\wp(\xi-\alpha+\beta))X Y+(\wp(\xi-\beta)\wp(\xi-\alpha+\beta)\nonumber \\
&-&t^2\wp(\xi+\beta)\wp(\xi+\alpha-\beta))X+(\wp(\xi+\alpha)\wp(\xi-\alpha+\beta)-t^2\wp(\xi-\alpha)\wp(\xi+\alpha-\beta))Y \nonumber \\
&+&(\wp(\xi+\alpha)\wp(\xi-\beta)-t^2\wp(\xi-\alpha)\wp(\xi+\beta))Z+t^2\wp(\xi-\alpha)\wp(\xi+\alpha-\beta)\wp(\xi+\beta)\nonumber \\
&-&\wp(\xi+\alpha)\wp(\xi-\beta)\wp(\xi-\alpha+\beta)\  ,
\end{eqnarray}
it is obvious that the first term of line 1 is equal to the corresponding term on the right hand-side of \eqref{3-legeq}. The rest of the equalities of the corresponding coefficients follow by
the same method as explained below. First, we make use of the Frobenius-Stickelberger formula \cite{FS} stated in  Appendix~\ref{appendix:b}, in terms of the variables $(\xi, \alpha, -\beta)$
\begin{align*}
\begin{vmatrix}
1 & \wp(\xi) & \wp'(\xi)\\
1 & \wp(\alpha) & \wp'(\alpha)\\
1 & \wp(-\beta) & \wp'(-\beta)
\end{vmatrix}
=2\frac{\sg(\xi+\alpha-\beta)\,\sg(\xi-\alpha)\,\sg(\alpha+\beta)\,\sg(\xi+\beta)}{\sg^3(\xi)\,\sg^3(\alpha)\,\sg^3(\beta)},
\end{align*}
and a similar relation with  $(\xi,-\alpha,\beta)$. If we divide the  former determinant by the latter one, we obtained the following expression for $t$ and $s$ in \eqref{eq:st}
\be
t=\frac{\wp'(\xi)(b-a)-A b-a B+\wp(\xi)(A+B)}{\wp'(\xi)(b-a)+Ab+aB-\wp(\xi)(A+B)},\; \: s=\frac{4(a-b)\wp'(\xi)}{(\wp'(\xi)(b-a)+Ab+aB-\wp(\xi)(A+B))^2}\ . \nn
\ee
Applying the elliptic addition formulae of the form, notably:
\begin{equation}\label{eq:addforms}
\wp(\xi)+\wp(\eta)+\wp(\xi\pm \eta)=\frac{1}{4}\left( \frac{\wp'(\xi)\mp \wp'(\eta)}
{\wp(\xi)-\wp(\eta)}\right)^2\    ,
\end{equation}
on \eqref{LHS}, we get on the one hand
\begin{eqnarray}\label{LHSS}
LHS&=&(1-t^2)X Y Z + (a+\wp(\xi)-\frac{(\wp'(\xi)-A)^2}{4(\wp(\xi)-a)^2}+t^2(-a-\wp(\xi)+\frac{(\wp'(\xi)+A)^2}{4(\wp(\xi)-a)^2}))Y Z \nn\\
&+& (b+\wp(\xi)-\frac{(\wp'(\xi)+B)^2}{4(\wp(\xi)-b)^2}+t^2(-b-\wp(\xi)+\frac{(\wp'(\xi-B)^2}{4(\wp(\xi)-b)^2}))X Z \nn\\
&+&(c+\wp(\xi)-\frac{(\wp'(\xi)-C)^2}{4(\wp(\xi)-c)^2}+t^2(-c-\wp(\xi)+\frac{(\wp'(\xi)+C)^2}{4(\wp(\xi)-c)^2}))X Y\nn\\
&+&((-a-\wp(\xi)+\frac{(\wp'(\xi)-A)^2}{4(\wp(\xi)-a)^2})(-b-\wp(\xi)+\frac{(\wp'(\xi)+B)^2}{4(\wp(\xi)-b)^2})\nn\\
&-&t^2((-a-\wp(\xi)+\frac{(\wp'(\xi)+A)^2}{4(\wp(\xi)-a)^2})(-b-\wp(\xi)+\frac{(\wp'(\xi)-B)^2}{4(\wp(\xi)-b)^2})))Z\nn\\
&+&((-a-\wp(\xi)+\frac{(\wp'(\xi)-A)^2}{4(\wp(\xi)-a)^2})(-c-\wp(\xi)+\frac{(\wp'(\xi)-C)^2}{4(\wp(\xi)-c)^2})\nn\\
&-&t^2((-a-\wp(\xi)+\frac{(\wp'(\xi)+A)^2}{4(\wp(\xi)-a)^2})(-c-\wp(\xi)+\frac{(\wp'(\xi)+C)^2}{4(\wp(\xi)-c)^2})))Y\nn\\
&+&((-b-\wp(\xi)+\frac{(\wp'(\xi)+B)^2}{4(\wp(\xi)-b)^2})(-c-\wp(\xi)+\frac{(\wp'(\xi)-C)^2}{4(\wp(\xi)-c)^2})\nn\\
&-&t^2((-b-\wp(\xi)+\frac{(\wp'(\xi)-B)^2}{4(\wp(\xi)-b)^2})(-c-\wp(\xi)+\frac{(\wp'(\xi)+C)^2}{4(\wp(\xi)-c)^2})))X\nn\\
&+&((a+\wp(\xi)-\frac{(\wp'(\xi)-A)^2}{4(\wp(\xi)-a)^2})(-b-\wp(\xi)+\frac{(\wp'(\xi)+B)^2}{4(\wp(\xi)-b)^2})(-c-\wp(\xi)\nn\\
&+&\frac{(\wp'(\xi)-C)^2}{4(\wp(\xi)-c)^2})+t^2((-a-\wp(\xi)+\frac{(\wp'(\xi)+A)^2}{4(\wp(\xi)-a)^2})(-b-\wp(\xi)\nn\\
&+&\frac{(\wp'(\xi)-B)^2}{4(\wp(\xi)-b)^2})(-c-\wp(\xi)+\frac{(\wp'(\xi)+C)^2}{4(\wp(\xi)-c)^2}))).
\end{eqnarray}
The proof is completed by using the relations \eqref{addition1} and subsequently \eqref{aA}, \eqref{eq:ABC} on the terms of \eqref{LHSS} and as well as on the right hand-side of \eqref{3-legeq}.

\section{ Frobenius-Stickelberger type identities}
\label{appendix:b}
\numberwithin{equation}{section}

Here we collect a number of results related to elliptic determinantal formulae of Frobenius and Frobenius-Stickelberger type (i.e. elliptic Cauchy and 
Vandermonde determinants). 
The Frobenius-Stickelberger formula, \cite{FS} is given by 
\begin{eqnarray}
&&\left| \begin{array}{ccccc} 1& \wp(x_1) &\wp'(x_1) &\cdots&\wp^{(n-2)}(x_1)\\
 1& \wp(x_2) &\wp'(x_2)&\cdots&\wp^{(n-2)}(x_2)\\
1& \wp(x_3) &\wp'(x_3)&\cdots&\wp^{(n-2)}(x_3)\\
\vdots&\vdots&\vdots&\ddots&\vdots\\
1&\wp(x_{n}) &\wp'(x_{n})&\cdots&\wp^{(n-2)}(x_{n})\\
\end{array}\right|\nn \\
&=&(-1)^{(n-1)(n-2)/2}\;1!2!3!...(n-1)!\;\frac{\sg(x_1+x_2+...+x_n)\prod_{i<j=1}^n \sg(x_i-x_j)}{\prod_{i=1}^n\sg^{n}(x_i)}\nn \\
\label{FSS}
\end{eqnarray}
Denoting the Frobenius-Stickelberger \textit{matrix} ${\mathcal P}(x_0,x_1,\dots,x_n)={\mathcal P}({\bs x})$ by: 
\be\label{eq:FSmatrix} 
{\mathcal P}({\bs x})=\left( \begin{array}{ccccc} 1& \wp(x_1) &\wp'(x_1) &\cdots&\wp^{(n-2)}(x_1)\\
 1& \wp(x_2) &\wp'(x_2)&\cdots&\wp^{(n-2)}(x_2)\\
1& \wp(x_3) &\wp'(x_3)&\cdots&\wp^{(n-2)}(x_3)\\
\vdots&\vdots&\vdots&\ddots&\vdots\\
1&\wp(x_{n}) &\wp'(x_{n})&\cdots&\wp^{(n-2)}(x_{n})\\
\end{array}\right)
\ee 
we have by using Cramer's rule the following factorisation formula:
\begin{equation}\label{eq:Frobinv}
\left[{\mathcal P}({\bs x})\cdot{\mathcal P}({\bs y})^{-1}\right]_{i,j}=
\frac{1}{\sg^{n}(x_i)}\,\Phi_\Sigma(x_i-y_j)\sg^{n}(y_j)\frac{\prod_{l=1}^n \sg(x_i-y_l)}{\prod_{l\neq j}
\sg(y_j-y_l)}\  ,
\end{equation}
in which ~$\Sigma\equiv\Sigma_{l=1}^n\,y_l$~. As a consequence we obtain from this the Frobenius determinantal 
formula, \cite{Frob}  
\be\label{eq:Frob}
{\rm det}\left( \Phi_\kp(x_i-y_j)\right)_{i,j=1,\dots,N}=\frac{\sg(\kp+\Sigma)}{\sg(\kp)}\,
\frac{\prod_{i<j}\,\sg(x_i-x_j)\,\sg(y_j-y_i)}{\prod_{i,j}\,\sg(x_i-y_j)}\ , \quad \Sigma:=\sum_{i=1}^N(x_i-y_i)\  .  
\ee  
Conversely, the Frobenius-Stickelberger formula \eqref{FSS} can be obtained from the Frobenius formula by 
a set of degenerate limits. The elliptic Lagrange interpolation formulae 
\begin{equation}
\prod_{i=1}^N \frac{\sigma(\xi - x_i)}{\sigma(\xi - y_i)}\,=\,
\sum_{i=1}^N \Phi_{-\Sigma}(\xi - y_i) 
\frac{\textstyle \prod_{j=1}^N \sigma(y_i - x_j)}{\textstyle 
\prod_{j=1\atop j\ne i}^N \sigma(y_i - y_j)}\   , 
\label{eq:Lagr1}   
\end{equation}
which holds if $\Sigma\neq 0$, and if $\Sigma=0$: 
\begin{equation}
\prod_{i=1}^N \frac{\sigma(\xi - x_i)}{\sigma(\xi - y_i)}\,=\,
\sum_{i=1}^N \left[ \zeta(\xi - y_i)  
- \zeta(x - y_i)\right]  
\frac{\textstyle \prod_{j=1}^N \sigma(y_i - x_j)}{\textstyle 
\prod_{j=1\atop j\ne i}^N \sigma(y_i - y_j)}\    , 
\label{eq:Lagr2}  
\end{equation}
where $x$ denotes any of the zeroes $x_i$, ($i=1,\dots,N$). Both \eqref{eq:Lagr1}  
can be obtained from the Frobenius formula \cite{Frob} by row-or column expansions (adding an extra row and column 
to the Frobenius matrix, say with $x_0=\xi$ and $y_0=\eta$, and then expanding along that row or column) and 
\eqref{eq:Lagr2} can subsequently be obtained from a limiting case of the latter.

\section{Proof of Equation \eqref{det:A} }
\label{appendix:c}
\numberwithin{equation}{section}

Here, we present the proof of the determinant in \eqref{det:A}. By definition of $A_{ilj}$ given in \eqref{eq:AB} we have 
\small
\begin{eqnarray}\label{eq:A_ilj}
&& \left| \begin{array}{cc} A_{ilj}& A_{il'j}\\
 A_{i'l j}&A_{i'l'j} \end{array}\right| =\frac{\sg(\wh{\xi}_l-\wh{\xi}_3)\sg(\wh{\xi}_{l'}-\wh{\xi}_3)}{S(\wh{\wt{\xi}}_i)\;S(\wh{\wt{\xi}}_{i'})\sg(\wh{\xi}_l-\xi_j-\beta)\sg(\wh{\xi}_{l'}-\xi_j-\beta)\sg^2(\wh{\xi}_3-\xi_j-\beta)} \nn \\
\nn \\
&&\hspace{2.5 cm} \Bigg[ \sg(\wh{\wt{\xi}}_i-\wh{\xi}_3-\wh{\xi}_l+\xi_j-\alpha+\beta)\sg(\wh{\wt{\xi}}_{i'}-\wh{\xi}_3-\wh{\xi}_{l'}+\xi_j-\alpha+\beta)\sg(\wh{\wt{\xi}}_i-\wh{\xi}_{l'}-\alpha)\sg(\wh{\wt{\xi}}_{i'}-\wh{\xi}_l-\alpha)\nn \\
&&\hspace{2.5 cm}-\sg(\wh{\wt{\xi}}_{i'}-\wh{\xi}_3-\wh{\xi}_l+\xi_j-\alpha+\beta)\sg(\wh{\wt{\xi}}_{i}-\wh{\xi}_3-\wh{\xi}_{l'}+\xi_j-\alpha+\beta)\sg(\wh{\wt{\xi}}_i-\wh{\xi}_l-\alpha)\sg(\wh{\wt{\xi}}_{i'}-\wh{\xi}_{l'}-\alpha)\Bigg],\nn \\
\end{eqnarray}
where
\begin{eqnarray*}
S(\xi)=\sg(\xi-\wh{\xi}_l-\alpha)\sg(\xi-\wh{\xi}_{l'}-\alpha)\sg(\xi-\wh{\xi}_3-\alpha).
\end{eqnarray*}
\normalsize
Noting that the difference in the bracket can be simplified by applying the three-term relation for the $\sg$-function in the following form:  
\begin{eqnarray}\label{eq:sigmaR}
\sg(x-a)\sg(y-b)\sg(z-b)\sg(w-a)&-&\sg(y-a)\sg(x-b)\sg(z-a)\sg(w-b)\nn\\
&=&\sg(z+y-a-b)\sg(x-y)\sg(x-z)\sg(b-a),\qquad \qquad
\end{eqnarray}
in which $x-y=z-w$. Making now the the following choice for  $x,\;y,\;z,\;w,\;a$ and $b$ in the identity \eqref{eq:sigmaR}: 
\begin{eqnarray*}
\begin{array}{ccc}
x=\wh{\wt{\xi}}_i-\wh{\xi}_3+\xi_j-\alpha+\beta&\: &y=\wh{\wt{\xi}}_{i'}-\wh{\xi}_3+\xi_j-\alpha+\beta \\
z=\wh{\wt{\xi}}_i-\alpha& \: &w=\wh{\wt{\xi}}_{i'}-\alpha \\
a=\wh{\xi}_l &\: & b=\wh{\xi}_{l'}
\end{array}
\end{eqnarray*}
the expression between brackets on the right-hand side of \eqref{eq:A_ilj} simplifies to 
\begin{eqnarray}\label{eq:bracket}
[\cdots]=\sg(-\wh{\xi}_3+\xi_j+\beta)\;\sg(\wh{\wt{\xi}}_i+\wh{\wt{\xi}}_{i'}-\wh{\xi}_l-\wh{\xi}_{l'}-\wh{\xi}_3+\xi_j-2\alpha+\beta)\;\sg(\wh{\wt{\xi}}_i-\wh{\wt{\xi}}_{i'})\;\sg(\wh{\xi}_{l'}-\wh{\xi}_l)\;.\qquad
\end{eqnarray}
Substituting the right-hand side of \eqref{eq:bracket} into \eqref{eq:A_ilj} and cancelling the first factor against the corresponding 
factor in the prefactor of \eqref{eq:A_ilj}, using the fact that $\sg$ is an odd function, we obtaine the desired result 
given by the determinant in \eqref{det:A}. In a similar way (or by making the obvious replacements $\alpha\leftrightarrow\beta$ and 
$\widetilde{\phantom{a}}\leftrightarrow\widehat{\phantom{a}}$~) the computation of the $2\times2$ determinant $B_{ilj}$ can be 
verified.


\end{document}